\documentclass[twocolumn,preprintnumbers,amsmath,amssymb,floatfix,showpacs,prl]{revtex4}
\usepackage{amsmath}
\usepackage{dcolumn}
\usepackage{braket}
\usepackage{graphicx}
\usepackage{hyperref}

\newcommand{\beq}{\begin{eqnarray}}
\newcommand{\eeq}{\end{eqnarray}}
\newcommand{\br}{{\bf r}}

\begin{document}
\title{Majorana Fermion Surface Code for Universal Quantum Computation}
\author{Sagar Vijay}
\author{Timothy H. Hsieh}
\author{Liang Fu}
\affiliation{Department of Physics, Massachusetts Institute of Technology, Cambridge, MA 02139, USA}
\begin{abstract}
We introduce an exactly solvable model of interacting Majorana fermions realizing $Z_{2}$ topological order with a $Z_{2}$ fermion parity grading and lattice symmetries permuting the three fundamental anyon types. We propose a concrete physical realization by utilizing quantum phase slips in an array of Josephson-coupled mesoscopic topological superconductors, which can be implemented in a wide range of solid state systems, including topological insulators, nanowires or two-dimensional electron gases, proximitized by $s$-wave superconductors.  Our model finds a natural application as a \emph{Majorana fermion surface code} for universal quantum computation, with a \emph{single-step} stabilizer measurement requiring no physical ancilla qubits, increased error tolerance, and simpler logical gates than a surface code with bosonic physical qubits. We thoroughly discuss protocols for stabilizer measurements, encoding and manipulating logical qubits, and gate implementations.
\end{abstract}
\maketitle

As originally proposed by Ettore Majorana, the Majorana fermion is a particle that is its own anti-particle \cite{Majorana}. In the condensed matter setting, Majorana fermions can emerge in topological superconductors as a special type of zero-energy, spatially-localized quasi-particle, 
formed by a coherent superposition of electron and hole excitations with equal amplitude \cite{Alicea, Beenakker_review}. 
Theory predicts that Majorana fermions can be created in a wide array of spin-orbit-coupled  
materials in proximity with conventional superconductors \cite{Fu_Kane, sarma, oreg, palee,chain}. 
Recently, the observation of zero-energy conductance peaks in such systems \cite{kouwenhoven, heiblum, yazdani, jia1} 
provides encouraging
hints of Majorana fermions \cite{law, potter}.  

Majorana fermions in topological superconductors are of great interest as they are predicted to exhibit exotic properties 
such as non-Abelian statistics \cite{moore-read, read-green, ivanov}, which have yet to be observed in nature. 
In addition to its theoretical significance, non-Abelian statistics provides the foundation for topological quantum computation, 
in which logical qubits are encoded in the topologically-degenerate states of non-Abelian anyons and qubit operations are performed by braiding \cite{Toric_Code}. 
Topological quantum computation has the theoretical advantage of being immune to errors caused by local perturbations \cite{Nayak-rmp}.  %
Demonstrating the non-Abelian statistics of Majorana fermions, however, requires braiding, fusing, and measuring the fusion outcome. 
This is a challenging task, as each of the above operations is yet to be experimentally achieved. Furthermore, braiding Majorana fermions alone is insufficient to perform the necessary gate operations for universal quantum computation.  
 
The ``surface code" \cite{Surf_Code_Kitaev, Surf_Code_Freedman} provides an alternative approach to universal quantum computation 
that uses measurements in an Abelian topological phase for gate operations and error correction. In the surface code, measurements of non-trivial commuting operators (stabilizers) are used to project onto a ``code state" and logical qubits are effectively encoded in the anyon charge of a region by ceasing certain stabilizer measurements  \cite{Raussendorf, Raussendorf_2, Fowler_Martinis, Fowler_Surface_Code}.  The logical gates necessary for universal quantum computation are realized through sequences of measurements used to move and braid the logical qubits.  An advantage of the surface code architecture is its remarkable ability for error detection and subsequent correction during qubit readout, as the nucleation of anyons through the action of a random operator can be reliably tracked through stabilizer measurements.  For a sufficiently low error rate per physical qubit measurement, scaling the size of the surface code produces an exponential suppression in propagated errors \cite{Error_Correction_Preskill}. Remarkably, recent experiments with superconducting quantum circuits have demonstrated the ability to perform high-fidelity physical gate operations and reliable error correction for a surface code of small size \cite{Martinis, Martinis_Surf_Code_Array, Steffen_Surf_Code_Array}.  

In this work, we introduce a new scheme for surface code quantum computation that uses Majorana fermions as the fundamental physical degrees of freedom and exploits their unique properties for encoding and manipulating logical qubits.  
Our surface code is based on a novel $Z_{2}$ topological order with fermion parity grading (defined below), which we demonstrate in 
a class of exactly solvable Hamiltonians of interacting Majorana fermions. 
We demonstrate that charging energy-induced quantum phase slips in superconducting arrays with Majorana fermions generate the required 
multi-fermion plaquette interactions, providing a physical realization of our model. We then describe a detailed physical implementation of the ``Majorana fermion surface code'', including physical qubit and stabilizer measurements, the creation of logical qubits, error correction, and logical gate operations required for universal quantum computation.

The Majorana fermion surface code poses significant benefits over a surface code with bosonic physical qubits. 
First, stabilizer measurements in the Majorana surface code can be performed in a \emph{single step}, whereas this requires several physical gate operations in the conventional surface code \cite{Fowler_Surface_Code, Fowler_Martinis}. As a result, we anticipate that the Majorana surface code has a significantly higher error tolerance.  Furthermore, our Majorana surface code operates with substantially less overhead, as it requires fewer physical qubits per encoded logical qubit, and uses no physical ancilla qubits. Second, we may tune the energy gap for anyon excitations in our physical realization of the Majorana plaquette Hamiltonian, increasing error suppression in the Majorana fermion surface code.  Finally, the lattice symmetries in the Majorana plaquette model permute the three fundamental anyon types, allowing a much simpler realization of the logical Hadamard gate. As we will show, the above advantages arise from the unique approach taken by our Majorana fermion surface code and the use of Majorana fermions as fundamental degrees of freedom. 
In particular, the unique property that a Majorana fermion is half of an ordinary fermion, with the consequence that two of them form a single physical qubit, is crucial to the Majorana fermion surface code. On the other hand, the non-Abelian statistics of Majorana fermions 
is of no relevance to our code, because it does not involve braiding them.     

This paper is organized as follows. First, we introduce a solvable model of interacting Majorana fermions on the honeycomb lattice realizing a novel ${Z}_2$ topological order with a ${Z}_{2}$ fermion parity grading and an exact $S_{3}$ anyon symmetry.  We propose a physical realization of this model, using charging energy in an array of mesoscopic superconductors \cite{Fu} to implement the required non-local interactions between multiple Majorana fermions. Next, we demonstrate that our model provides a natural setting for the Majorana fermion surface code, in which a logical qubit is encoded in a set of physical qubits formed from Majorana fermions.  We present a physical implementation of the Majorana surface code and propose detailed protocols for performing gate operations for universal quantum computation.      

\section{I. Majorana Plaquette Model}
\begin{figure}
$\begin{array}{cc}
\includegraphics[trim = 120 60 120 110, clip = true, width=0.2\textwidth, angle = 0.]{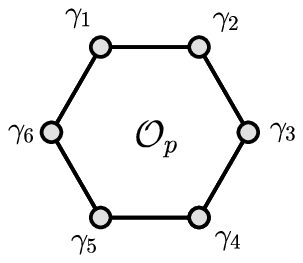} &
\includegraphics[trim = 50 90 130 10, clip = true, width=0.21\textwidth, angle = 0.]{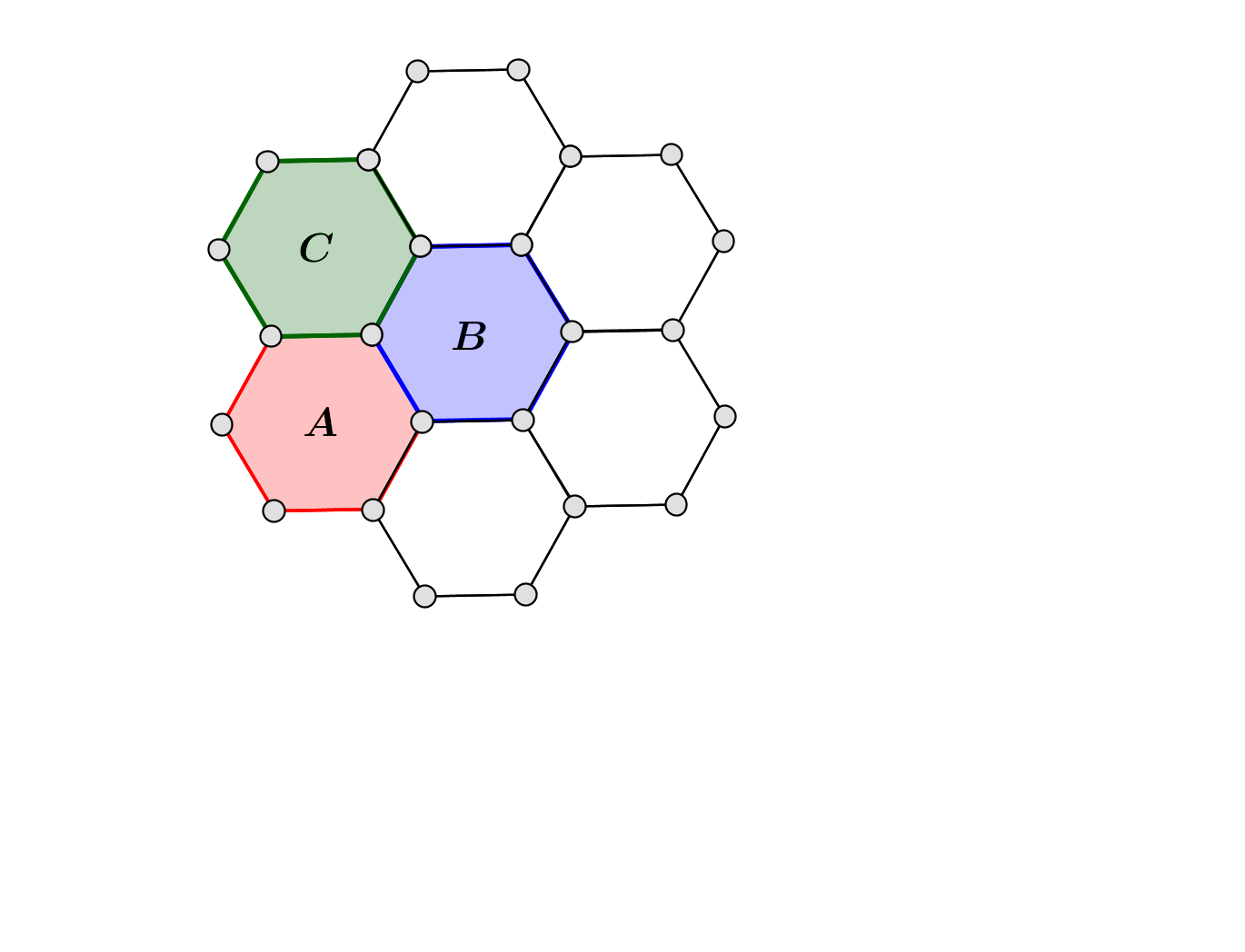}\\
\text{(a)} & \text{(b)}
\end{array}$
\caption{We consider a honeycomb lattice with (a) a single Majorana fermion on each lattice site, so that the $\mathcal{O}_{p}$ operator is the product of the six Majorana fermions on the vertices of a hexagonal plaquette. The colored plaquettes in (b) correspond to the three distinct bosonic excitations that may be obtained by violating a plaquette constraint. \label{fig:Plaquette_1}}
\end{figure}

 We begin by considering a honeycomb lattice with one Majorana fermion ($\gamma$) on each lattice site; the Majorana fermions satisfy canonical anti-commutation relations $\{\gamma_{n},\gamma_{m}\} = 2\delta_{nm}$. The Hamiltonian is defined as the sum of operators acting on each hexagonal plaquette:
\begin{align}\label{eq:Hamiltonian}
H = -u\sum_{p}\mathcal{O}_{p} \hspace{.25in} \mathcal{O}_{p} \equiv i\prod_{n \in\rm{vertex} (p)}\gamma_{n}. 
\end{align}
We note that this model was mentioned in a work by Bravyi, Terhal and Leemhuis \cite{bravyi}, although its novel topological order and anyon excitations were not studied there. 
It suffices to consider $u>0$ below, as the case of $u<0$ can be mapped to $u>0$ by changing the sign of the Majorana fermions on one sublattice.  
The operator $\mathcal{O}_{p}$ is the product of the six Majorana fermions on the vertices of plaquette $p$ as shown in Figure \ref{fig:Plaquette_1}a.  Since any two plaquettes on the honeycomb lattice share an even number of vertices, all of the plaquette operators commute, and the ground-state $\ket{\Psi_{0}}$ is defined by the condition 
\beq
\mathcal{O}_{p}\ket{\Psi_{0}} = \ket{\Psi_{0}},  \label{p}
\eeq 
for all plaquettes $p$. We note that quite generally, Hamiltonians of interacting Majorana fermions with commuting terms may be realized on any lattice, so long as any pair of operators in the Hamiltonian only has overlapping support over an even number of Majorana fermions.

We demonstrate  that the above Majorana plaquette model (\ref{eq:Hamiltonian}) realizes a ${Z}_{2}$ topological order of Fermi systems by considering the ground-state degeneracy and elementary excitations. 
 First, we place the system on a torus by imposing periodic boundary conditions, and find a four-fold degenerate ground-state by counting the number of degrees of freedom and constraints on the full Hilbert space.   
For an $N$-site honeycomb lattice, the $2^{N/2}$-dimensional Hilbert space of Majorana fermions 
is constrained by the fixed total fermion parity:
\begin{align}
\Gamma \equiv i^{N/2}\prod_{n}\gamma_{n}
\end{align}
For convenience, we choose a unit cell for the honeycomb lattice consisting of three plaquettes labeled $A, B$ and $C$, as shown in Figure \ref{fig:Plaquette_1}b. We observe that on the torus, the product of plaquette operators on each of the $A$, $B$ and $C$-type plaquettes is equal to the total fermion parity: 
\begin{align}
\Gamma = \prod_{p\in A}\mathcal{O}_{p} = \prod_{p\in B}\mathcal{O}_{p} = \prod_{p\in C}\mathcal{O}_{p}
\end{align}
The operators $\{\mathcal{O}_{p}\}$ on any one type of plaquette fix one-third of the plaquette eigenvalues via the condition (\ref{p}), and 
impose $2^{N/6 - 1}$ constraints on the Hilbert space.  The number of unconstrained degrees of freedom is therefore given by:
\begin{align}
D = 2^{\frac{N}{2} - 1}/\left( 2^{\frac{N}{6} - 1}\right)^{3} = 4, 
\end{align}
which yields a four-fold ground state degeneracy for the Majorana plaquette model on the torus. 

\begin{figure}
\includegraphics[trim = 80 130 80 15, clip = true, width=0.31\textwidth, angle = 0.]{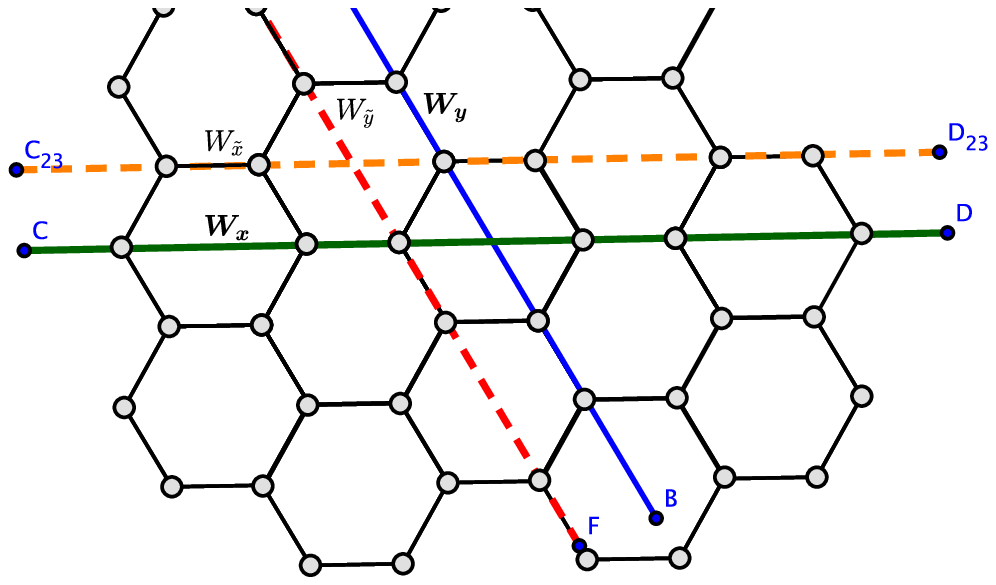} 
\caption{The action of the commuting Wilson loop operators $W_{x}$ and $W_{y}$ is shown above as the product of the Majorana fermions on the lattice sites intersected by the appropriate colored lines. The operator $w_{x}$  anti-commutes with $W_{x}$ and takes the ground-state between two topological sectors.\label{fig:Wilson_Loop}}
\end{figure}

The ground state degeneracy is of a topological nature, as the four ground-states are distinguished only by non-local operators. To see this, we construct a Wilson loop operator $W_\ell$, defined as a product of Majorana bilinears on a non-contractible loop $\ell$ on the torus:
\begin{align}\label{eq:Wilson_Loop}
W_\ell \equiv \prod_{n,m\in\ell}(i\gamma_{n}\gamma_{m}). 
\end{align}
such that $W_{\ell}^{2} = 1$, so that the Wilson loop has eigenvalues $\pm 1$. Consider the operators $W_x$ and $W_y$ on the two non-trivial cycles of the torus $\ell_x$ and $\ell_y$, as shown in Figure \ref{fig:Wilson_Loop}. 
Since $\ell_{x}$ and $\ell_{y}$ traverse an even number of vertices over any plaquette and do not contain any common lattice sites, we have 
$[W_{x},W_{y}] = [W_{x}, H] = [W_{y}, H] = 0$. 
Furthermore, we may construct Wilson loop operators $W_{\tilde x}$ and $W_{\tilde y}$ on loops $\tilde{\ell}_x$ and $\tilde{\ell}_y$, where $\tilde{\ell}_x$ is shifted from $\ell_x$ by a basis vector parallel to $\ell_y$ and likewise for $\tilde{\ell}_y$, such that $\{W_{\tilde{x}}, W_y\} = \{W_{\tilde{y}},W_x\} = 0$.      
As before, $W_{\tilde{x}}$ and $W_{\tilde{y}}$ commute with each other and with the Hamiltonian. Therefore, the four degenerate ground-states may be distinguished by their eigenvalues under $W_x$ and $W_y$, with $W_{\tilde{x}}$ and $W_{\tilde{y}}$ transforming the ground-states between distinct sectors. In analogy with conventional $Z_2$ gauge theory, we may identify the Wilson loop operators $W_{x,y}$ with electric charges traversing the torus in two different directions, 
and $W_{\tilde{x}, \tilde{y}}$ as magnetic fluxes on a dual lattice.    

Gapped excitations above the ground state are obtained by flipping the eigenvalue of $\mathcal{O}_{p}$ from $+1$ to $-1$ on one or more plaquettes.  Since the total fermion parity is fixed and equal to the product of all plaquette operators of each type, plaquette eigenvalues can only be flipped on pairs of plaquettes of the \emph{same} type. This is achieved by string operators of the form (\ref{eq:Wilson_Loop}), now acting on open paths and anti-commuting with the plaquette operators at the two ends of the path, thereby creating a pair of anyon excitations.

An important feature of our Majorana plaquette model, the conservation of total fermion parity---a universal property of Fermi systems---makes it impossible to create or annihilate two excitations living on different types of plaquettes, or change one type of plaquette excitation into another.  As a result, there are three distinct elementary plaquette excitations, labeled $A$, $B$ and $C$, by plaquette type. 
To determine their statistics, we braid these excitations by acting with Majorana hopping operators $i\gamma_{n}\gamma_{m}$ on lattice bonds \cite{Supp_Mat}.  We find that all three types of plaquette excitations have boson self-statistics and mutual semion statistics, i.e., braiding two distinct plaquette excitations generates a quantized Berry phase of $\pi$.  From the elementary plaquette excitations we may build composite excitations $AB$, $BC$, $AC$ and $ABC$ by flipping the eigenvalues of the $\mathcal{O}_{p}$'s on two or three adjacent plaquettes. Among these, the composite excitation $ABC$ is simply a physical Majorana fermion, since the Majorana operator $\gamma_n$ acting on a lattice site flips the eigenvalues of the $\mathcal{O}_{p}$'s on the three surrounding $A$, $B$ and $C$ plaquettes.   
In contrast, the composite excitations $AB$, $BC$, $AC$ are anyons, with fermion self-statistics and mutual semion statistics with the elementary excitations. We call these excitations composite Majorana fermions, as they are created by a string of physical Majorana fermions.  
A summary of the braiding statistics for all anyons in our Majorana plaquette model  is given in the following table:
\begin{align*}
  \begin{tabular}{|c|c|c|c|c|c|c|c|c|}
  \hline
    & $1$ & $A$ & $B$ & $C$ & $AB$ & $BC$ & $AC$ & $ABC$\\
    \hline
$1$ & $+1$ & $+1$ & $+1$ & $+1$ & $+1$ & $+1$ & $+1$ & $+1$\\ 
    \hline
$A$ & $+1$ & $+1$ & $-1$ & $-1$ & $-1$ & $+1$ & $-1$ & $+1$\\
    \hline  
$B$ & $+1$ & $-1$ & $+1$ & $-1$ & $-1$ & $-1$ & $+1$ & $+1$\\
    \hline
$C$ & $+1$ & $-1$ & $-1$ & $+1$ & $+1$ & $-1$ & $-1$ & $+1$\\
    \hline
$AB$ & $+1$ & $-1$ & $-1$ & $+1$ & $-1$ & $-1$ & $-1$ & $+1$\\
    \hline
$BC$ & $+1$ & $+1$ & $-1$ & $-1$ & $-1$ & $-1$ & $-1$ & $+1$\\
    \hline
$AC$ & $+1$ & $-1$ & $+1$ & $-1$ & $-1$ & $-1$ & $-1$ & $+1$\\
    \hline
$ABC$ & $+1$ & $+1$ & $+1$ & $+1$ & $+1$ & $+1$ & $+1$ & $-1$\\
    \hline
  \end{tabular}
\end{align*}

Strange as it may appear, the existence of eight types of quasiparticle excitations is a generic property of $Z_2$ topologically ordered phases in Fermi systems, due to the conservation of fermion parity. 
Consider artificially dividing the above quasiparticles into two groups: $(1, A, B, AB)$ and $(ABC, BC, AC, C)=ABC\times (1, A, B, AB)$. 
The former is equivalent to the four quasi-particles in $Z_2$ gauge theory coupled to a bosonic Ising matter field, as realized in Kitaev's toric code \cite{Toric_Code} or Wen's plaquette model \cite{Wen_Plaquette}.    
The latter group of quasiparticles is obtained by attaching a physical Majorana fermion to the former. The conservation of total fermion parity guarantees that the two groups of quasi-particles cannot transform into each other in a closed system, and thus have separate identities. We refer to the presence of two groups of excitations with different fermion parity as a $Z_2$ fermion parity grading.   

A remarkable property of the Majorana plaquette model is that crystal symmetries of the honeycomb lattice permute the three fundamental anyon excitations, $A, B$ and $C$, by interchanging the three types of plaquettes.    
Examples of such lattice symmetries include $\pi/3$ rotations about the center of a plaquette, 
and translation by any primitive lattice vector. 
These symmetries of the honeycomb lattice provide a microscopic realization of the $S_3$ anyon symmetry that permutes quasiparticle sectors, as recently studied in the abstract formalism of topological field theory by considering the symmetries of the $K$-matrices of Abelian topological states \cite{Teo, Kitaev}. 

\section{II. Physical Realization}
\subsection{A. Physical Platforms}
In this section, we show that the Majorana plaquette model can be physically realized in an array of mesoscopic topological superconductors that are Josephson coupled.  
A wide range of material platforms for engineering a topological superconductor have been proposed and are being experimentally studied \cite{Alicea, Beenakker_review}. 
As it will be clear in the following, the scheme we propose for realizing the Majorana plaquette model is independent of which platform is used. 
For the sake of concreteness, we use a platform based on topological insulators in describing the general scheme below, and discuss other platforms based on nanowires and two-dimensional electron gas with spin-orbit coupling in section II.C.   

We place a array of hexagon-shaped $s$-wave superconducting islands on a topological insulator (TI) to induce a superconducting proximity effect on the TI surface states. The Hamiltonian for this superconductor-TI hybrid system is given by
\beq
H_{0} &=& \int d \br (-i v)\psi^\dagger(\br) \left(\partial_x s_y -\partial_y s_x - \mu\right) \psi(\br)  \nonumber \\
&+&  \sum_j \int d \br_j \left[ \Delta  e^{i \varphi_j}   \psi^\dagger_\uparrow(\br_j) \psi^\dagger_\downarrow (\br_j) + \mathrm{h.c.} \right], \label{h0}
\eeq
where $\psi=(\psi_\uparrow, \psi_\downarrow)^T$ is a two-component fermion field and $s_{x,y}$ are spin Pauli matrices. 
The first term describes the pristine TI surface states, with a single spin-non-degenerate Fermi surface and helical spin texture in momentum space. The second term 
describes the superconducting proximity effect: $\br_j$ belongs to the region underneath the $j$-th superconducting island, whose phase is denoted by $\varphi_j$.     

\begin{figure}
\includegraphics[trim = 0 65 140 0, clip = true, width=0.25\textwidth, angle = 0.]{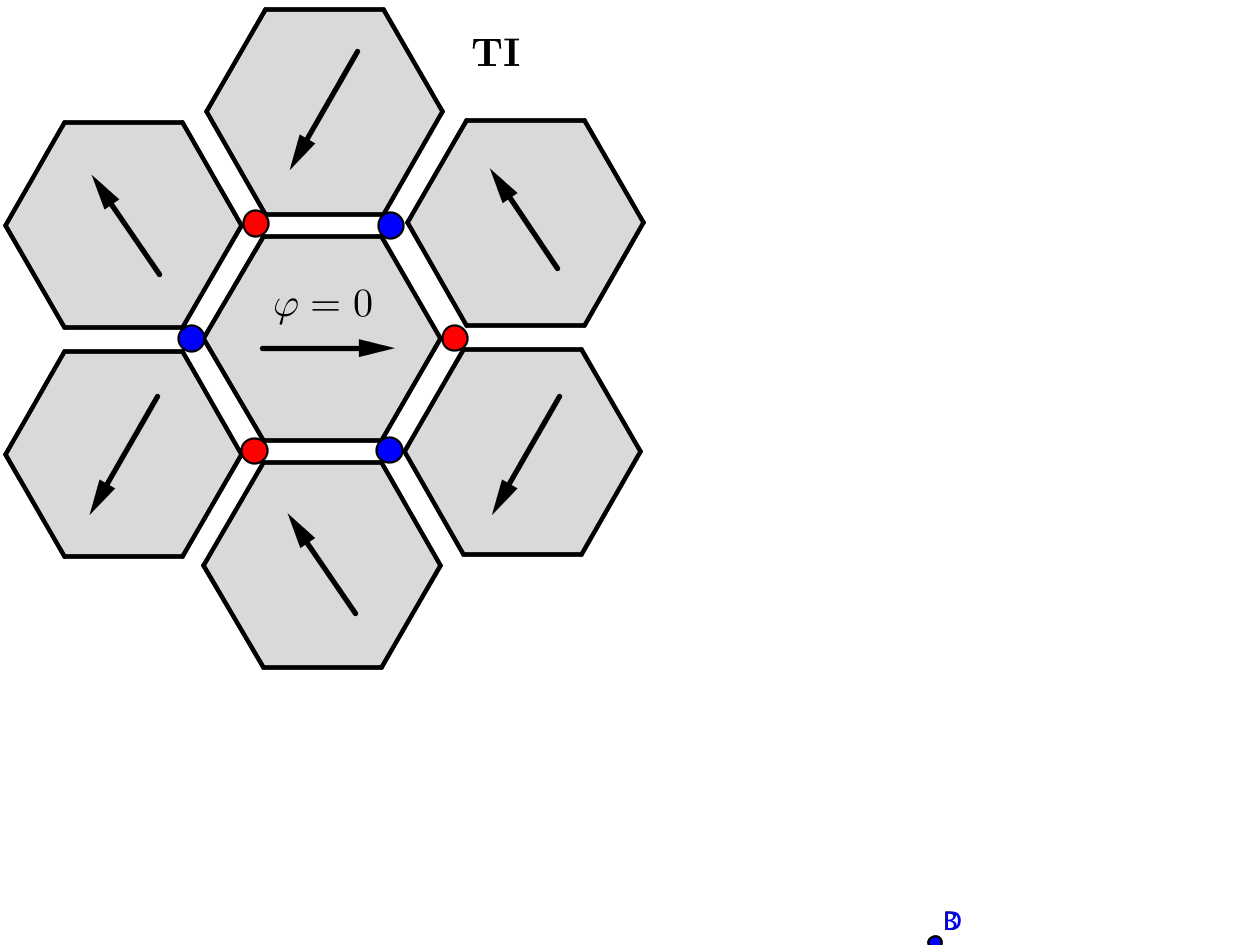}
\caption{Array of hexagonal $s$-wave superconducting islands placed on a TI surface. Each arrow points in the direction of the relative phase of the associated island, with $\varphi = 0$, $\pm 2\pi/3$. This produces a  honeycomb lattice of vortices (blue) and anti-vortices (red) at tri-junctions, hosting Majorana fermions.\label{fig:Phase_Slip}}
\end{figure}

As found by Fu and Kane \cite{Fu_Kane}, a vortex or anti-vortex trapped at a tri-junction, where three islands meet, hosts a single Majorana fermion zero mode.  
Let us consider setting up the phases of superconducting islands to realize an array of vortices and anti-vortices at tri-junctions. 
For example, the phases can be set to $\varphi_j=  0, 2\pi/3$ and $-2\pi/3$ on the $A$, $B$ and $C$-type islands respectively,  as shown in Fig. \ref{fig:Phase_Slip}. 
This yields a 2D array of Majorana fermions on a honeycomb lattice. In practice, the desired phase configuration can be engineered by external electrical circuits \cite{private} and/or magnetic flux. 
Alternatively, applying a perpendicular magnetic field generates a vortex lattice. 
These vortices may naturally sit at these tri-junctions where the induced superconductivity is weak, leading to the desired lattice of Majorana fermions.

We take the size of the islands to be larger than the coherence length of the superconducting TI surface states. Under this condition, 
Majorana fermions at different sites have negligible wavefunction overlap, preventing any unwanted direct coupling between them. 
(We note that even weak couplings from wavefunction overlap will not affect the $Z_{2}$ topological order of the Majorana plaquette model, due to its finite  
energy gap.)
Nonetheless, as we show below, the charging energy of superconductors induces a nonlocal interaction  between 
the six Majorana fermions on each island, providing the key ingredient of the Majorana plaquette model.   

\subsection{B. Phase-Slip Induced Multi-Fermion Interactions}
The important but subtle interplay between Majorana fermions and charging energy was first recognized by Fu and formulated for superconductors with a fixed number of electrons \cite{Fu}. Later works have extended it to multiple superconductors connected by Josephson coupling and single-electron tunneling \cite{XuFu,  VanHeck1, Hutzen, VanHeck2}.  
In all of these cases, the charging energy of a given superconductor induces quantum phase slips  $\varphi \rightarrow \varphi \pm 2\pi$, from which   
the Majorana fermions in the superconductor 
acquire a minus sign: $\gamma_i \rightarrow -\gamma_i$. This property is  
due to the inherently double-valued dependence of Majorana operators on the superconducting phase \cite{Fu}. 
  
 In our setup for the Majorana plaquette model,  the charging energy of the superconducting islands exerts even more dramatic and interesting effects on the Majorana fermions at tri-junctions, which have not been previously studied. In the presence of a charging energy, the phase of each island becomes a quantum rotor. The kinetic energy of the rotor is provided by the charging energy $E_c$, which depends on the capacitance between an island and the rest of the array, and is described by the following Hamiltonian
\beq
H_{c} &= 4E_{c} \displaystyle\sum_j (\hat{n}_j - n_{g})^{2}  \label{ec}
\eeq
where 
$
\hat{n}_{j}\equiv (-i)\cdot {\partial}/{\partial \varphi_{j}}
$
is the Cooper pair number operator for the $j^{\mathrm{th}}$ island and $n_g$ is the offset charge, which can be tuned by an externally applied electric field.  The potential energy of the rotor is provided by the Josephson coupling $E_J$ between adjacent superconducting islands, given by 
\beq
H_J = -E_{J} \sum_{\langle j,j'\rangle} \cos (\hat{\varphi}_j   -\hat{\varphi}_{j'} - a_{jj'} ), \label{ej}
\eeq
where $a_{jj'} = \varphi_{0, j} - \varphi_{0,j'}$ is externally set up such that the minimum of the Josephson energy corresponds to  $\varphi_j = \varphi_{0,j} \mod 2\pi$, 
with $\varphi_{0, j} = 0, 2\pi/3$ and $-2\pi/3$ for the $A$, $B$ and $C$-type islands, respectively.

\begin{figure*}
$\begin{array}{cccc}
\includegraphics[trim = 0 34 140 0, clip = true, width=0.21\textwidth, angle = 0.]{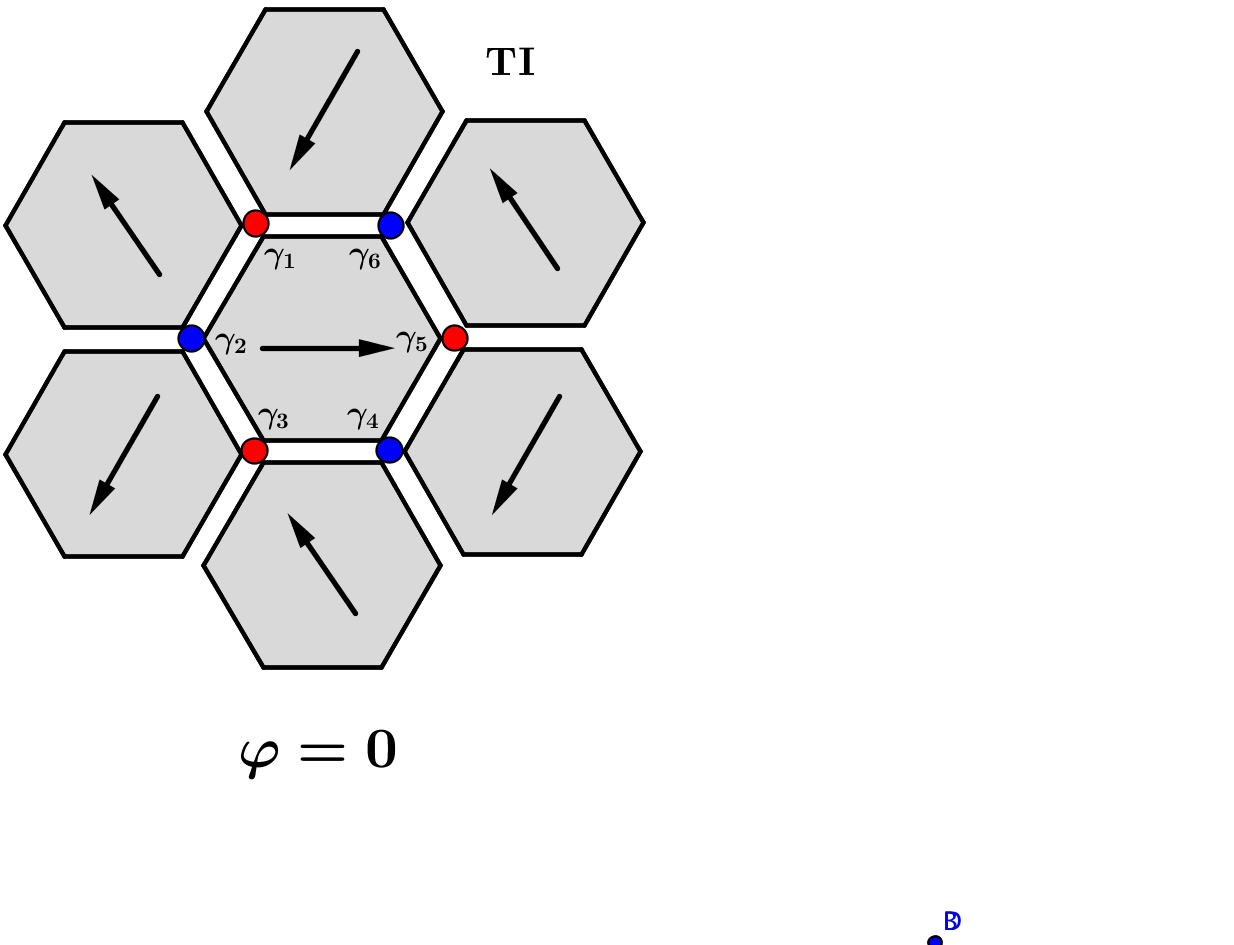} & 
\includegraphics[trim = 0 34 140 0, clip = true, width=0.21\textwidth, angle = 0.]{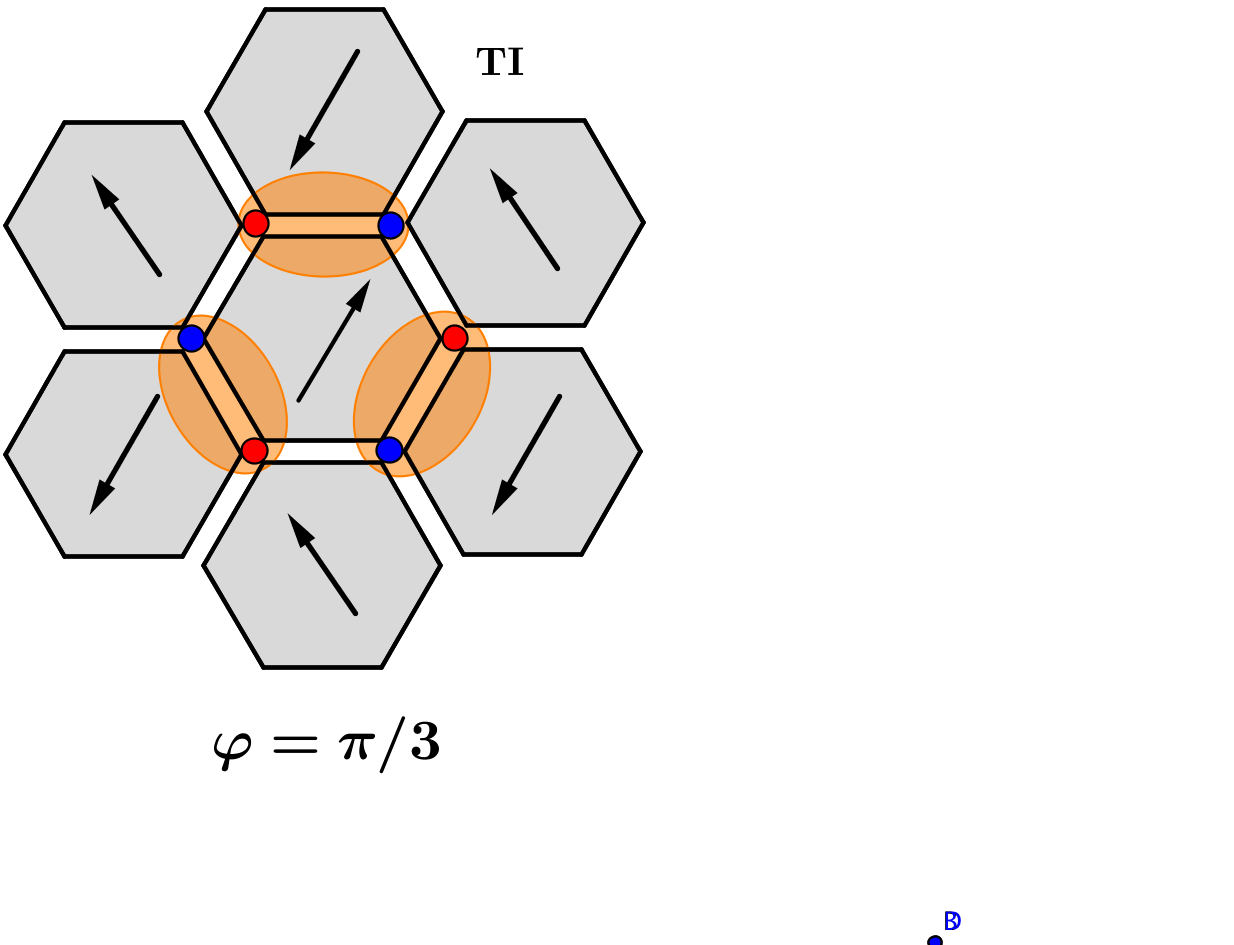} &
\includegraphics[trim = 0 34 140 0, clip = true, width=0.21\textwidth, angle = 0.]{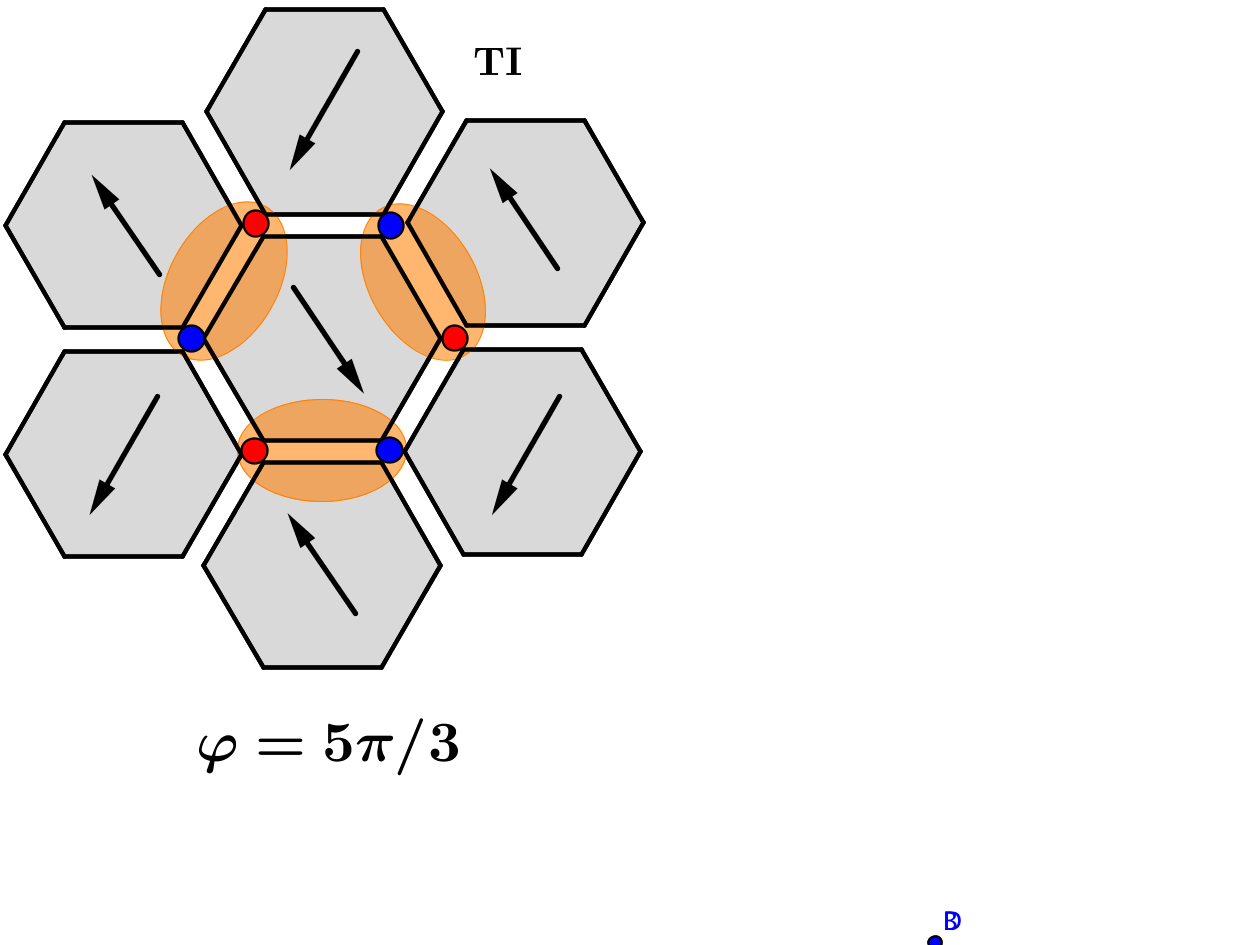} & 
\includegraphics[trim = 0 34 140 0, clip = true, width=0.21\textwidth, angle = 0.]{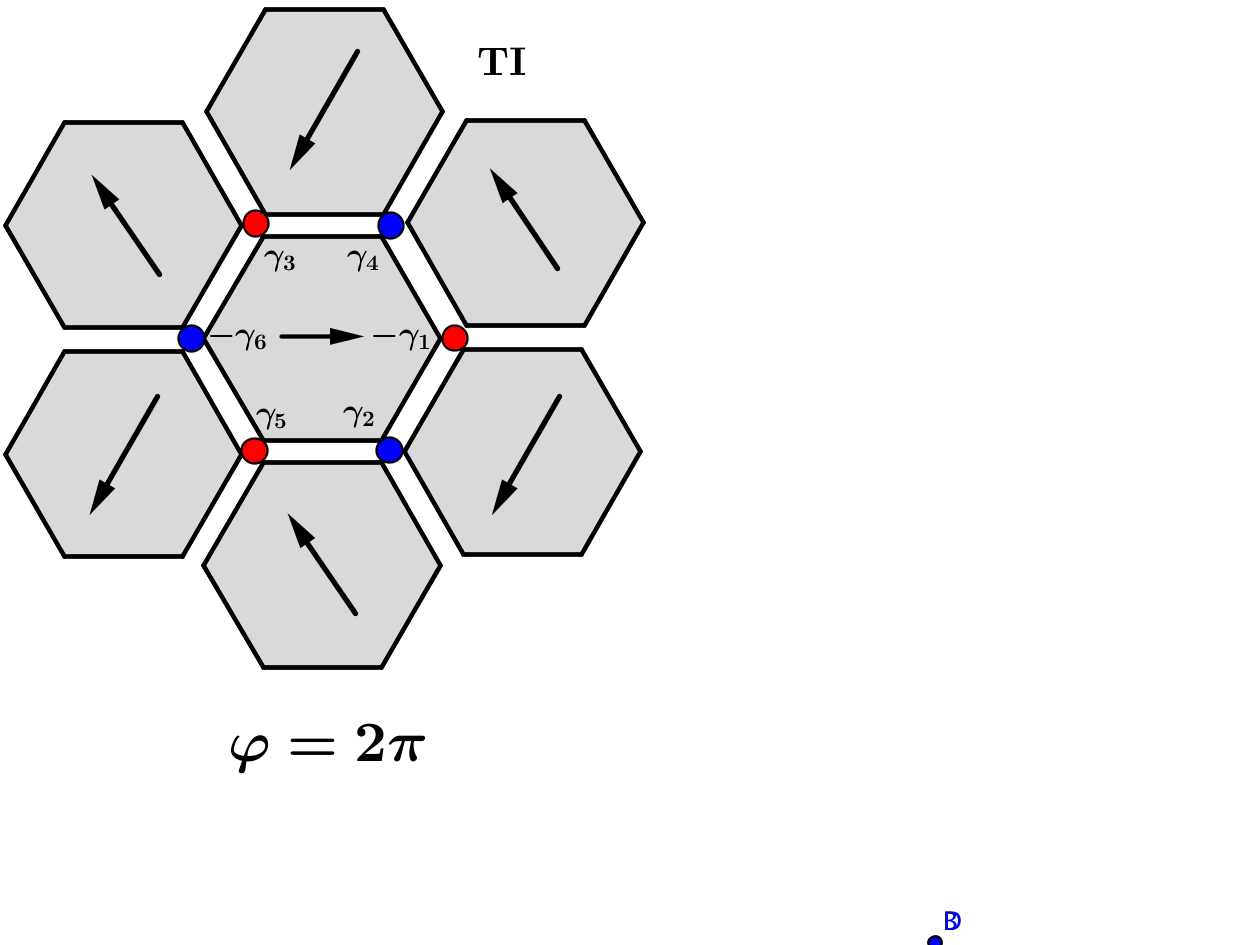}\\
\text{(a)} & \text{(b)} & \text{(c)} & \text{(d)}
\end{array}$
\caption{Schematic of a $2\pi$ phase-slip on the central superconducting island in a hexagonal superconducting array on a TI surface, with the phase of the central island indicated in each panel.  When the phase difference between neighboring islands is $\pi$, the pair of Majorana fermions on the shared edges couple \cite{Fu_Kane} as indicated.  The $2\pi$ phase-slip permutes the Majorana fermions as shown, leading to the transformation in (\ref{u1}).\label{fig:Phase_Slip_Panel}}
\end{figure*}

Combining (\ref{h0}), (\ref{ec}) and (\ref{ej}), the full Hamiltonian for our setup, i.e. an array of superconducting islands on a TI surface, is given by  
\beq \label{eq:Total_H}
H = H_0 + H_c + H_J. 
\eeq
We work in the regime $E_J \gg E_c$.
Under this condition, low-energy states of the quantum rotor on a given island  $\varphi_j$ consist of small-amplitude fluctuations around each potential minimum $\varphi_{0,j} + 2\pi m$. Moreover, different minima are connected by quantum phase slips, in which the phase $\varphi$ tunnels through a high energy barrier to wind  by $2\pi n$, with $n$ an integer. The small-amplitude phase fluctuations around a potential minimum correspond to a quantum harmonic oscillator, and thus generate a set of energy levels given by   
\beq
\epsilon^0_{\alpha} \approx (\alpha +1/2) \sqrt{8 E_J E_c}. \label{ep0}
\eeq   
with $\alpha\in\mathbb{N}$.

On the other hand, quantum phase slips on a superconducting island strongly couple to the Majorana fermions that reside on the border with its neighbors, previously obtained by holding the phase fixed at $\varphi_{0,j}$. In other words, Majorana fermions enter the low-energy effective theory of (\ref{eq:Total_H}) via quantum phase slips induced by small the charging energy on each superconducting island. This new physics makes our system different from a conventional Cooper pair box.  
Remarkably, the action of a quantum phase slip involves Majorana fermions in a way that 
depends periodically on the phase winding number $n \mod 6$. 
Consider, for example, phase slips at the central superconducting island in Fig. 3. 
For $n=1$, a $2\pi$ phase slip $\varphi =0 \rightarrow 2\pi$ cyclically permutes the three Majorana fermions bound to vortices in the counterclockwise direction, 
and the three Majorana fermions bound to anti-vortices in the clockwise direction, i.e., 
\beq
&\varphi=0 \rightarrow 2\pi:&  \gamma_1 \rightarrow \gamma_3, \gamma_3 \rightarrow \gamma_5, \gamma_5 \rightarrow -\gamma_1 \nonumber \\
& &  \gamma_2 \rightarrow -\gamma_6, \gamma_4 \rightarrow \gamma_2, \gamma_6 \rightarrow \gamma_4, \label{u1}
\eeq
as shown in Figure \ref{fig:Phase_Slip_Panel}, where $i=1,...,6$ labels the six Majorana fermions at vertices of this island in clockwise order.
The physical movement of Majorana fermions induced by phase slips is a unique and attractive advantage of our setup, compared to other setups in which the 
positions of Majorana fermions are fixed \cite{XuFu, VanHeck1, Hutzen, VanHeck2}. 
On the other hand, for $n=3$, a $6\pi$ phase slip takes each Majorana fermion over a full circle and back to its original position, from which it acquires a minus sign \cite{Fu_Kane}, i.e., 
\beq
\varphi=0 \rightarrow 6\pi: \; \gamma_i \rightarrow -\gamma_i.  \label{u3}
\eeq      
Only for $n=6$ does each Majorana fermion come back to its original position unchanged. 

We now add up the contributions of various phase slips to derive an effective Hamiltonian for Majorana fermions as a function of the offset charge $n_g$ for each state of the harmonic oscillator:   
\beq
H_\alpha (n_g) =\epsilon^0_\alpha +  \sum_{n=1}^6  ( t_{\alpha, n} \hat{U}_{n} e^{i 2\pi n n_g} +  \mathrm{h.c.} ). \label{hc}
\eeq 
Here $\epsilon^0_\alpha$ is the quantized energy of the harmonic oscillator given by (\ref{ep0}), which is the same for all internal states of the Majorana fermions. The second term describes quantum phase slips: $t_{\alpha,n}$ denotes the amplitude of the $\alpha$-th energy level of the harmonic oscillator tunneling 
between two potential minima that differ by $2\pi n$, while $\hat{U}_{n}$ is the unitary operator acting on the Majorana fermions due to a $2\pi n$ phase slip. The coupling $t_{\alpha, n}$ depends on the energy barrier in the phase slip event and can be modulated by tuning $E_c/E_J$; for example, $t_{\alpha, 1} \propto e^{-\sqrt{8 E_{J}/E_{c}}}$ \cite{Yale_Group}.
The offset charge $n_g$ provides an Aharonov-Bohm flux proportional to the winding number $n$. 

The Hamiltonian (\ref{hc})  is analogous to the Bloch Hamiltonian 
that describes the band structure of a particle hopping in a one-dimensional periodic potential, with the offset charge $n_g$ playing the role of crystal momentum.    
Importantly, the phase particle carries internal degrees of freedom resulting from Majorana fermions $\gamma_1, ..., \gamma_6$ that are unique to our system.   
A phase slip that moves the phase particle to a different potential minimum also permutes the Majorana fermions as shown in (\ref{u1},\ref{u3}), similar to a spinful particle hopping in the presence of a non-Abelian gauge field.  
These permutations are represented by the unitary operators $\hat{U}_n$ in the effective Hamiltonian (\ref{hc}) acting  on Majorana fermions. 
For example, the operator $\hat{U}_1$ that generates the transformation (\ref{u1})  is given by: 
\beq
\hat{U}_1 &=&   \frac{1+ \gamma_2 \gamma_3}{\sqrt{2}} \cdot \frac{1+ \gamma_4 \gamma_5}{\sqrt{2}} \cdot \frac{1- \gamma_6\gamma_1}{\sqrt{2}} \nonumber \\
&\times & \frac{1+ \gamma_1 \gamma_2}{\sqrt{2}} \cdot \frac{1+ \gamma_3 \gamma_4}{\sqrt{2}} \cdot \frac{1+ \gamma_5 \gamma_6}{\sqrt{2}}. 
\eeq
It follows from the addition of phase slips that $\hat{U}_n = (\hat{U}_1)^n$. In particular, the unitary operator $\hat{U}_3$, which takes $\gamma_i$ to $-\gamma_i$ as shown in (\ref{u3}), 
has a simple form: 
\beq
\hat{U}_3= -\prod_{i=1}^6 \gamma_i = i  \mathcal{O},  
\eeq
where $\mathcal{O}$ is the plaquette operator defined in the Majorana plaquette model (\ref{p}). On the other hand, for $n = 1,2, 4$ or $5$, $U_n$ is a sum of operators $\gamma_i \gamma_j$, $\gamma_i \gamma_j \gamma_k \gamma_l$ and $i \mathcal{O}$.   

Substituting the expressions for the $U_n$'s into (\ref{hc}), we find that the effective Hamiltonian induced by the small charging energy of a single island takes the following form
\beq\label{eq:H_eff}
H_{\alpha} (n_g) = \epsilon^0_\alpha + \Delta_\alpha(n_g) \mathcal{O} + V_\alpha(n_g),   
\eeq
with 
\beq
\Delta_\alpha(n_g) = \sum_{m=1}^5  {t}_{\alpha, m} \sin(2\pi m n_g). \label{delta}
\eeq
 $V_\alpha(n_g)$ includes a constant $t_{\alpha,6} \cos (12\pi n_g)$, as well as Majorana bilinear and quartic operators 
generated by  phase slips with winding number $n \neq 0 \mod 3$.  
Unlike $\mathcal{O}$, these operators on neighboring islands do not commute. 
From now on, we assume that $V$ can be treated as a perturbation to the Majorana plaquette model that does not destroy the $Z_2$ topological order of the gapped phase.    
An alternative setup without the presence of $V$ will be presented in a forthcoming work \cite{Future_Work}.

\subsection{C. Discussion}
In deriving the effective Hamiltonian (\ref{hc}), we have implicitly assumed that 
Majorana fermions are the only low-lying excitations involved in phase slip events, separated by an energy gap from other Andreev bound states in the junctions between islands. This assumption is valid because of the finite size of the islands, which leads to a discrete Andreev bound-state spectrum with a finite gap for all values of the phase. 
The presence of this gap justifies our derivation of the effective Hamiltonian (\ref{hc}) in a controlled manner.    

Over the last few years, considerable experimental progress has been made in hybrid TI-superconducting systems. 
Proximity-induced superconductivity and supercurrents have been observed in a number of TI materials \cite{brinkman, stanford, iop, molenkamp, moler, mason, harlingen}. Low-temperature scanning tunneling microscopy (STM) experiments have found proximity-induced superconducting gap on TI surface states, and the tunneling spectrum of Abrikosov vortices shows a zero-bias conductance peak, which is robust in a range of magnetic field and splits at higher field \cite{jia1}. This peak has been attributed to the predicted Majorana fermion zero-modes in the vortex cores of superconducting TI surface states. 
In view of these rapid, unabated advances,  we regard 
the hybrid TI-superconductor system as a very promising material platform for realizing the Majorana plaquette model and 
studying the exciting physics of Majorana fermions enabled by quantum phase slips. 
  
Besides TIs, a two-dimensional electron gas (2DEG) with spin-orbit coupling (such as InAs) can be driven into a helical state with an odd number of spin-polarized Fermi surfaces by an external Zeeman field, which provides another promising platform for realizing topological superconductivity via proximity effect \cite{SDS-PRL, Alicea-PRB}. 
In this topological regime, vortices and tri-junctions of a superconducting 2DEG host a single Majorana fermion, similar to the TI surface.  
Thus our proposed setup for the Majorana plaquette model in Section IIA directly applies to this system as well.

In addition to TIs and 2DEG, 
(quasi-)one-dimensional semiconductors and metals with strong spin-orbit coupling have become a hotly pursued system for Majorana fermions \cite{sarma, oreg, palee}. 
Signatures of Majorana fermions were reported in 2012, based on 
the observation of zero-bias conductance peak in hybrid nanowire-superconductor systems \cite{kouwenhoven,heiblum}.    
One can envision a network of nanowires in proximity with Cooper-pair boxes to realize our Majorana plaquette model. 
In this direction it is worth noting that a new physical system---a nanowire with an epitaxially grown superconductor layer---
has been recently introduced to study Andreev bound-states in the presence of charging energy \cite{marcus}.

Many other physical systems for Majorana fermions have been theoretically proposed and experimentally pursued, too numerous to list.  
Regardless of the particular system, 
non-local interactions between multiple Majorana fermions emerge from the charging energy of superconductors via quantum phase slips, 
and in the universal regime, such interactions are determined by the transformation of Majorana fermions under phase slips, as we have shown in Section IIB.   

Finally, we note several previous works related to our Majorana plaquette model and its physical realization. 
In Ref. \cite{XuFu}, Xu and Fu first introduced a model of interacting Majorana fermions that realizes $Z_2$ topological order. 
This model involves 4-body and 8-body plaquette interactions on square and octagonal plaquettes in a two-dimensional lattice. Physical realizations of this model were proposed using an array of superconductor islands in proximity with either 2D TI \cite{XuFu}, or semiconductor nanowires \cite{divincenzo}.   
The 4-body nonlocal interaction between Majorana fermions comes directly from the charging energy, whereas the 8-body interaction comes from a {\it high-order} ring-exchange process generated by single-electron tunneling between islands. 
In comparison, our Majorana plaquette model on the honeycomb lattice has the theoretical novelty of possessing an exact anyon permutation symmetry,  
and can be realized in a much simpler manner using an array of superconductors on a 3D TI with global phase coherence, 
with all  the required interactions coming directly from the charging energy. We also note a recent work on 
lattice models of Majorana fermions in Abrikosov vortices on a superconducting TI surface \cite{franz}, which use different interactions and 
do not exhibit topological order.

\section{III. Majorana Surface Code}

In the rest of this work, we demonstrate that the Majorana plaquette model finds a natural application as a ``Majorana fermion surface code", on which universal quantum computation and error correction may be performed. The main idea of the surface code is to (i) use anyons of the Majorana plaquette model to encode logical qubits, (ii) manipulate anyons to perform gate operations on logical qubits, and (iii) use commuting measurements of the Majorana plaquette operators for error correction.  
We will describe the detailed implementation of the Majorana surface code, including the creation of logical qubits, error correction, and protocols for logical gate operations required for universal quantum computation. 

The surface code architecture \cite{Fowler_Martinis, Surf_Code_Kitaev, Surf_Code_Freedman} is a \emph{measurement-based} scheme for quantum computation. 
It uses projective measurements of commuting operators---called ``stabilizers''---acting on a 2D array of physical qubits to produce a highly-entangled ``code state" $\ket{\psi}$. Logical qubits are created by stopping the measurement of certain commuting operators to create ``holes''. 
The different possible anyon charges at a hole are the degrees of freedom that define a logical qubit. Logical gates are realized by manipulating and braiding holes via a sequence of measurements. 

A key advantage of the surface code is its remarkable capability for \emph{error detection}. The random measurement of an operator in the surface code corresponds to nucleating pairs of anyons, a process that can be reliably measured by tracking the eigenvalues of the commuting stabilizers. 
Reliable error detection hinges on (i) having a large number of physical qubits for a given encoded logical qubit, and (ii) a sufficiently low error rate for stabilizer measurements \cite{Fowler_Martinis}.  For the previously studied surface code with bosonic physical qubits, it has been estimated \cite{Fowler_Error_1, Fowler_Error_2} that below a threshold as high as $\sim1\%$ error-rate per physical qubit operation, scaling the size of the surface code permits an exponential suppression of errors propagated. 
This error tolerance makes the surface code architecture one of the most realistic approaches to practical, large-scale quantum computation. 

Recent practical realizations of the surface code have used superconducting qubits coupled to a microwave transmission line resonator to perform qubit manipulations and measurements \cite{Martinis, Martinis_Surf_Code_Array, Steffen_Surf_Code_Array}. Here, a {physical} qubit is defined by two energy levels arising from quantization of number/phase fluctuations in a conventional Cooper pair box.    
The surface code is implemented on a 2D array of physical qubits with the four-qubit interactions of Kitaev's toric code Hamiltonian \cite{Toric_Code} as the set of commuting stabilizers.  The four-qubit stabilizer is measured by performing a sequence of single and two-qubit gates between the four \emph{physical} qubits and additional ancilla qubits \cite{Fowler_Martinis}.
Experiments have demonstrated the remarkable ability to operate these physical gates with fidelity above the threshold required for surface code error correction \cite{Martinis}.  Recent experiments have also used error detection to preserve entangled code states on a surface code with a 9$\times$1 \cite{Martinis_Surf_Code_Array} and a 2$\times$2 \cite{Steffen_Surf_Code_Array} array of stabilizers. It remains to be shown that logical qubits can be successfully encoded and manipulated via logical gates in these surface code arrays.  

\subsection{A. Implementation}
We implement the Majorana surface code on a 2D array of Majorana fermions by performing projective measurements of the Majorana plaquette operators $\{\mathcal{O}_p\}$, 
which form a complete set of commuting stabilizers. For the remainder of this paper, we will use `plaquette operators' and 'stabilizers' interchangeably to refer to $\{\mathcal{O}_{p}\}$. A practical {\it physical} system for implementing the Majorana surface code is the superconductor-TI hybrid system introduced in the previous section. We place a array of superconducting islands on the TI surface, which are strongly Josephson coupled. By introducing external circuits or applying fluxes, we engineer  
the Josephson coupling between islands to achieve the phase configuration in Figure \ref{fig:Phase_Slip}, leading to a honeycomb lattice of Majorana fermions at tri-junctions.

\begin{figure}
$\begin{array}{c}
\hspace{.2in}\includegraphics[trim = 474 344 285 254, clip = true, width=0.34\textwidth, angle = 0.]{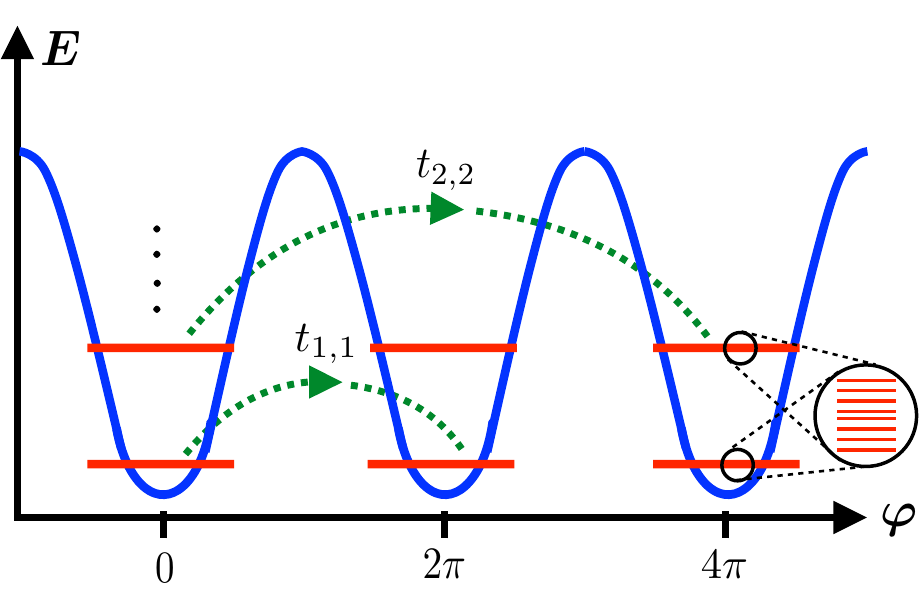} \\
\text{(a)}\\
\includegraphics[trim = 295 317 355 154, clip = true, width=0.36\textwidth, angle = 0.]{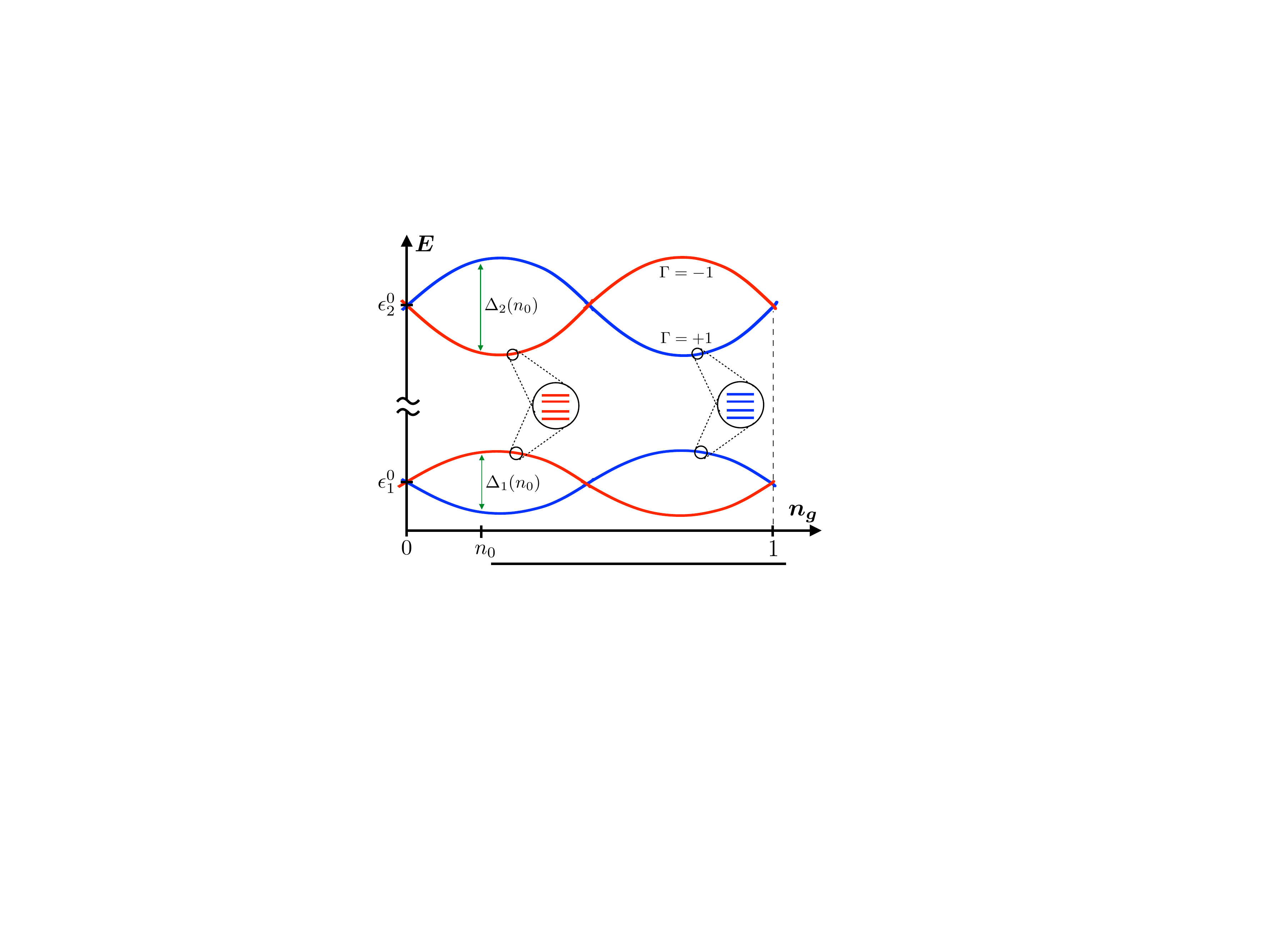}\\
\text{(b)}
\end{array}$
\caption{(a.) Schematic of the harmonic oscillator energy levels of the effective Hamiltonian (\ref{eq:H_eff}), centered at $\varphi = 2\pi n$, with the $2\pi$ and $4\pi$ phase-slip amplitudes for the lowest energy levels shown.  In (b), we show a schematic plot of the two lowest harmonic oscillator levels as a function of the gate-charge. The energy splittings $\Delta_{1}$ and $\Delta_{2}$ are between states with even ($\Gamma = +1$) and odd fermion parity ($\Gamma = -1$) within the first and second harmonic oscillator levels, respectively. Each level within a fixed fermion parity sector is nearly four-fold degenerate.\label{fig:Fermion_Parity}}
\end{figure}

To perform a projective measurement of the Majorana plaquette operator on a given island, i.e., a single stabilizer, we 
decrease the Josephson coupling of the island with the rest of the array to activate quantum phase slips from the small but non-zero charging energy on this island. As shown by the effective Hamiltonian in (\ref{eq:H_eff}), these quantum phase slips (partially) lift the degeneracy between states in the eight-dimensional Fock space of the six Majorana fermions.  
In particular, for every energy level of the harmonic oscillator, there is an energy splitting $\Delta_\alpha(n_g)$ between states of Majorana fermions with 
$\Gamma = +1$ (even fermion parity) and with $\Gamma = -1$ (odd fermion parity) from (\ref{delta}),  where $\Gamma$ is the stabilizer eigenvalue; this is shown schematically in Figure \ref{fig:Fermion_Parity}b.
Therefore, the charging energy of the island creates an energy difference between different stabilizer eigenstates.  Furthermore, the \emph{energy gap} between the two lowest harmonic oscillator levels on the island is a function of the stabilizer eigenvalue $\Gamma = \pm 1$, and in the limit of negligible interaction $V$ takes the following form: 
\begin{align}
\Delta E_{\Gamma}(n_{g}) = \epsilon_{0} + \left[\Delta_{2}(n_{g}) - \Delta_{1}(n_{g})\right]\Gamma + \ldots
\end{align}
where $\epsilon_{0} \equiv \epsilon^{0}_{2} - \epsilon^{0}_{1} \approx \sqrt{8E_{J}E_{c}}$. The sensitivity of the energy gap to the stabilizer eigenvalue now permits a stabilizer measurement by simply measuring the energy gap. By shining a probe microwave beam on this island, we may measure the phase shift of the transmitted photons to determine the gap between the two harmonic oscillator levels \cite{Top_Transmon, Schoelkopf}.  

We now perform these stabilizer measurements on all of the superconducting islands to project onto an eigenstate of the Majorana plaquette Hamiltonian (\ref{eq:Hamiltonian}); this will be our reference ``code state". We continue to perform measurements on all hexagonal islands in each cycle of the surface code in order to maintain the state. In subsequent cycles, we may encode logical qubits into the code state and manipulate the qubits via measurement.  While projection onto the code state and error correction in the surface code rely exclusively on measuring the six-Majorana plaquette interaction, manipulation of logical qubits also requires measuring nearest-neighbor Majorana bilinears on the hexagonal lattice.  This may be done by tuning the phase of neighboring superconducting islands to bring the pair of Majorana fermions on the shared edge sufficiently close together \cite{Fu_Kane}, so that the resulting wavefunction overlap further splits the nearly four-fold degeneracy within a single fermion parity sector shown in Figure \ref{fig:Fermion_Parity}b.  Again, the Majorana bilinear may be measured by shining a probe beam to measure the energy gap to the next harmonic oscillator level. 

Using the commuting six-Majorana operators in our plaquette model to realize a surface code provides unique advantages over the more conventional surface code with bosonic physical qubits.  First, while a four-spin stabilizer measurement in the usual surface code requires performing 6$-$8 gates/measurements between a set of physical and ancilla qubits \cite{Fowler_Martinis, Fowler_Error_1}, stabilizer eigenvalues in the Majorana surface code are obtained via a \emph{single-step measurement} by shining a probe beam.  
We emphasize that even when measurement is not being performed, the intrinsic charging energy of the islands generates a finite gap $\Delta_{1}(n_{g})$ to creating anyon excitations, and naturally \emph{suppresses} errors at temperatures $k_{B}T < \Delta_{1}(n_{g})$.
We anticipate that the corresponding error tolerance for scalable quantum computation is substantially improved for the Majorana surface code. 
Second, the Majorana surface code operates with lower overhead than its bosonic counterpart, using three-qubit stabilizers, and requiring no ancilla qubits.  Finally, the anyon transmutation required to perform a logical Hadamard gate in the conventional surface code corresponds to a duality transformation that exchanges the star and plaquette toric code operators.  This operation is quite difficult to perform on a single logical qubit as it also requires lattice surgery to patch the transformed logical qubit back into the remaining surface code \cite{Fowler_Martinis, Fowler_Hadamard}. As lattice symmetries permute anyon sectors in the Majorana plaquette model, anyon transmutation in the Majorana surface code corresponds to a \emph{lattice translation} of the logical qubit, substantially simplifying the Hadamard gate implementation.

\subsection{B. Logical qubits and error correction}
\begin{figure}
$\begin{array}{c}
\includegraphics[trim = 61 110 47 10, clip = true, width=0.34\textwidth, angle = 0.]{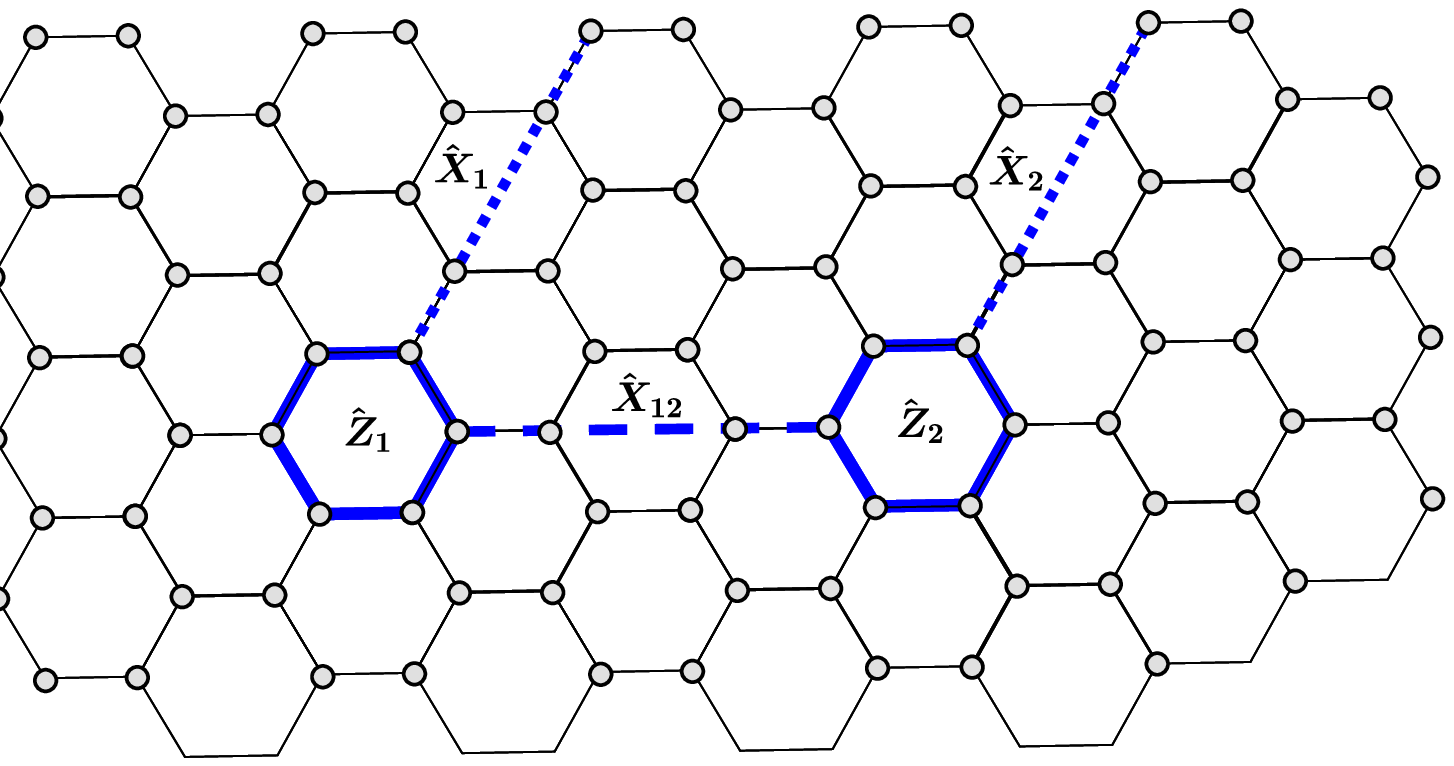}\\
\text{(a)}\\
\\
\includegraphics[trim = 21 83 83 29, clip = true, width=0.3\textwidth, angle = 0.]{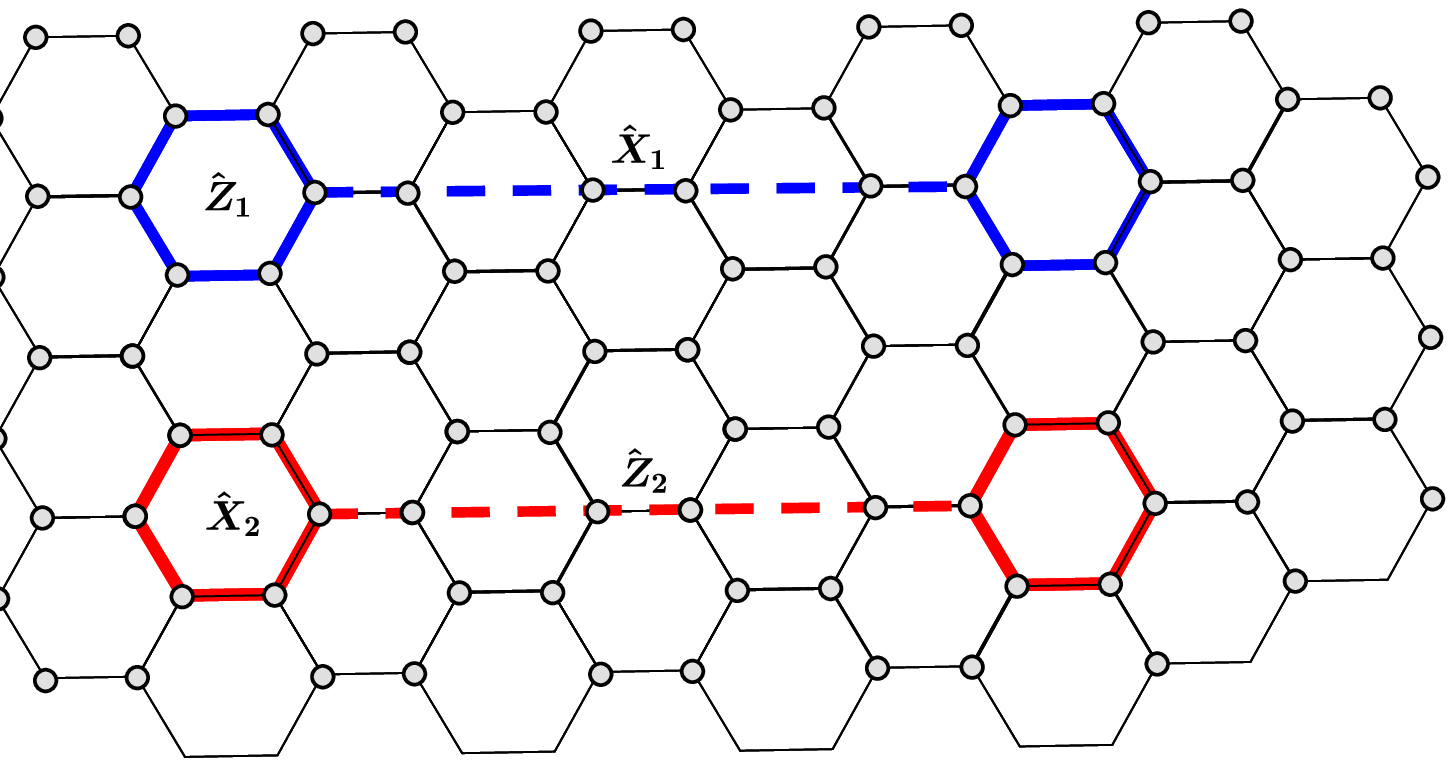}\\
\text{(b)}\\
\\
\includegraphics[trim = 22 60 164 50, clip = true, width=0.31\textwidth, angle = 0.]{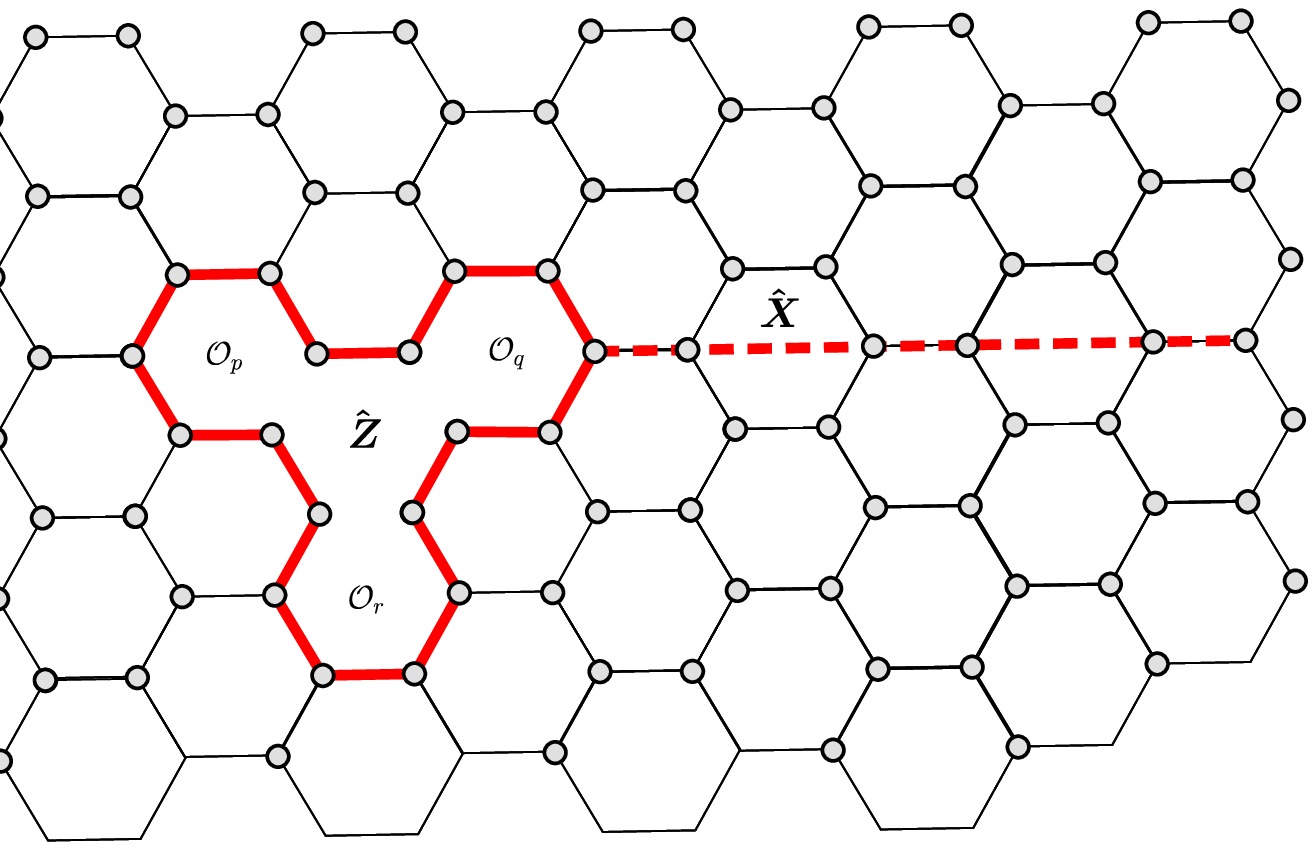}\\
\text{(c)}
\end{array}$
\caption{Logical qubits in the Majorana surface code. In (a) we stop the measurement of two plaquette operators in subsequent surface code cycles, increasing the ground-state degeneracy by a factor of four. If we take $\hat{Z}_{1}$ and $\hat{Z}_{2}$ to be the logical $\hat{Z}$ operators for the two encoded  qubits, the corresponding $\hat{X}_{1}$ and $\hat{X}_{2}$ operators are given by Wilson lines connecting to the boundary. The two qubits may be coherently manipulated by applying the operator $\hat{X}_{12}$ as shown. In practice, it is simpler to define logical qubits by stopping the measurement of pairs of plaquettes of a single type, with the logical $\hat{X}$ and $\hat{Z}$ defined as shown in (b). We may also consider a logical qubit made of several `holes', as in (c), to minimize errors during qubit manipulation.\label{fig:Logical_Qubit}}
\end{figure}

Logical qubits may be encoded in the surface code by ceasing the measurement of the plaquette operator on a hexagonal superconducting island in a surface code cycle, while continuing measurements on all other plaquettes. In theory, we could stop measuring a single plaquette and define a two-level system, with the $\hat{Z}$ and $\hat{X}$ operators of the logical qubit defined by the plaquette operator and a Wilson line connecting the plaquette to the boundary, respectively. A pair of such qubits on the $A$-type plaquettes is shown in Figure \ref{fig:Logical_Qubit}a, where the solid and dashed lines correspond to products of Majorana fermions that define the indicated logical operators. The two qubits shown may also be coherently manipulated by acting with the Wilson line operator connecting the two plaquettes, denoted $\hat{X}_{12}$. 

In practice, however, it is difficult to manipulate qubits with an operator that connects to a distant boundary, so it is simpler to encode a logical qubit by stopping the stabilizer measurement on two well-separated plaquettes of the same type. We choose to only manipulate two of the four resulting degrees of freedom by defining $\hat{Z} \equiv \mathcal{O}_{p}$ and $\hat{X} \equiv W_{pq}$, the Wilson line operator connecting the two plaquettes. We use the opposite convention to define the logical $\hat{Z}$ and $\hat{X}$ operators for a qubit on the adjacent $B$ plaquettes; an example of such logical qubits is shown in Figure \ref{fig:Logical_Qubit}b. We note that when such a qubit is created, it is automatically initialized to an eigenstate of the plaquette operator, with eigenvalue given by the measurement performed in the previous surface code cycle.  As a result, logical qubits of type $A$ ($B$) are initialized to an eigenstate of the $\hat{Z}$ ($\hat{X}$) logical operator.

To reduce errors during qubit manipulation, we may define a qubit by ceasing measurement of multiple adjacent plaquettes as shown in Figure \ref{fig:Logical_Qubit}b. In this particular case, the logical operator $\hat{X}$ is still a Wilson line connecting to another set of distant `holes'. However, the logical $Z$ is defined as $\hat{Z} \equiv \mathcal{O}_{p}\otimes\mathcal{O}_{q}\otimes\mathcal{O}_{r}$. For the remainder of our discussion, we will consider logical qubits with only a single plaquette operator used to define the logical $\hat{Z}$; the generalization to larger qubits is straightforward. 

Errors may occur during qubit manipulation, including (1) single-qubit errors due to the unintended measurement of a local operator involving an even number of Majorana fermions and (2) measurement errors.  Single-qubit error correction may be performed on logical qubits by constantly measuring the remaining plaquette eigenvalues during surface code cycles.  Since only \emph{pairs} of plaquettes may be flipped simultaneously by a random measurement, corresponding to the nucleation of a pair of anyons of a single type, detecting the change of an odd number of plaquette eigenvalues in a single surface code cycle will generally signal the presence of a random measurement performed on a nearby logical qubit. More precisely, when a stabilizer eigenvalue changes in a surface code cycle, it is efficient to store the location of that stabilizer, and wait several code cycles, accumulating a spacetime diagram of stabilizer errors as additional errors occur \cite{Fowler_Error_1, Fowler_Error_2, Error_Correction_Preskill}.  After sufficiently many code cycles, the spacetime diagram may be used to determine the most likely configuration of Wilson lines that could have generated those errors \cite{Fowler_Surface_Code, Fowler_Martinis} using a minimum-weight perfect matching algorithm \cite{Minimum_Weight_Algorithm, Fowler_Error_3}. Errors may be subsequently corrected by software when performing logical qubit manipulations and readouts \cite{Fowler_Martinis}. Random measurement errors involve incorrectly registering the eigenvalue of a plaquette operator; these are naturally corrected by performing multiple surface code cycles to verify the accuracy of a measurement.

\subsection{C. Logical gate implementations}
The Majorana surface code may be used for universal quantum computation by implementing CNOT, $T$, and Hadamard gates on logical qubits; this has been extensively studied in the context of the surface code architecture with underlying bosonic degrees of freedom \cite{Fowler_Hadamard, Fowler_Martinis}.  Here, we describe the implementations of these gates in our realization of quantum computation with a Majorana surface code.  Our gate implementations follow the spirit of the implementations presented in \cite{Fowler_Martinis}.

All gates in the Majorana surface code are implemented on logical qubits via a sequence of measurements. Let $\hat{U}$ be the desired unitary we wish to perform on the quantum state of several logical qubits, defined by the logical operators $\{\hat{X}_{i}\}$ and $\{\hat{Z}_{i}\}$. It is convenient to keep track of the transformation of the logical state by monitoring the transformation of logical operators $\hat{X}_{i}\rightarrow \hat{U}\hat{X}_{i}\hat{U}^{\dagger}$, $\hat{Z}_{i}\rightarrow \hat{U}\hat{Z}_{i}\hat{U}^{\dagger}$. In practice, performing the appropriate sequence of measurements will yield the transformation $W$, such that:
\begin{align}
&\hat{U}\hat{X}_{i}\hat{U}^{\dagger} = \pm \hat{W}\hat{X}_{i}\hat{W}^{\dagger}\\
&\hat{U}\hat{Z}_{i}\hat{U}^{\dagger} = \pm \hat{W}\hat{Z}_{i}\hat{W}^{\dagger}
\end{align}
where the signs depend on the outcomes of specific measurements performed. These measurement outcomes are stored in a software and used to correctly interpret the readout of a logical qubit.

In what follows, we will often demonstrate our gate implementations in an ``operator picture", where a set of operators in the surface code $\hat{o}_{1}$, $\ldots$, $\hat{o}_{n}$ and $\hat{p}_{1}$, $\ldots$, $\hat{p}_{m}$ with eigenvalues $\pm 1$ are measured in an appropriate sequence. This implements a logical gate via the desired transformations:
\begin{align}
&\hat{Z} \rightarrow \hat{Z}\otimes\prod_{j=1}^{n}\hat{o}_{j}= \hat{U}\hat{Z}\hat{U}^{\dagger}\\
&\hat{X} \rightarrow \hat{X}\otimes\prod_{j=1}^{m}\hat{p}_{j} = \hat{U}\hat{X}\hat{U}^{\dagger}.
\end{align}
In practice, the measured outcomes for the $\{\hat{o}_{i}\}$ and $\{\hat{p}_{j}\}$ operators will be stored by software and used to obtain the above transformations during logical qubit readout.

{\bf CNOT gate:}  A CNOT gate takes two qubits -- a ``control" and a ``target" -- and flips the value of the target qubit based on the value of the control, and returns the control unchanged. The action of a CNOT takes the following form in the basis of two-qubit states:
\begin{align}
\hat{C} = \left(
\begin{array}{cccc}
1 & 0 & 0 & 0\\ 
0 & 1 & 0 & 0\\
0 & 0 & 0 & 1\\
0 & 0 & 1 & 0
\end{array}
\right)
\end{align}
A CNOT gate may be implemented by braiding logical qubits in the Majorana surface code. In the simplest case, a CNOT between two logical qubits of different types is implemented through a single braiding operation that produces an overall sign if the hexagonal ends of both qubits contain an anyon, due to the $\pi$ mutual statistics, demonstrated in Section I. 
In the following section, we first describe the procedure to move a logical qubit along a given type of plaquette before discussing the braiding procedure required to produce a CNOT gate.

\begin{figure}
$\begin{array}{c}
\includegraphics[trim = 15 210 180 60, clip = true, width=0.4\textwidth, angle = 0.]{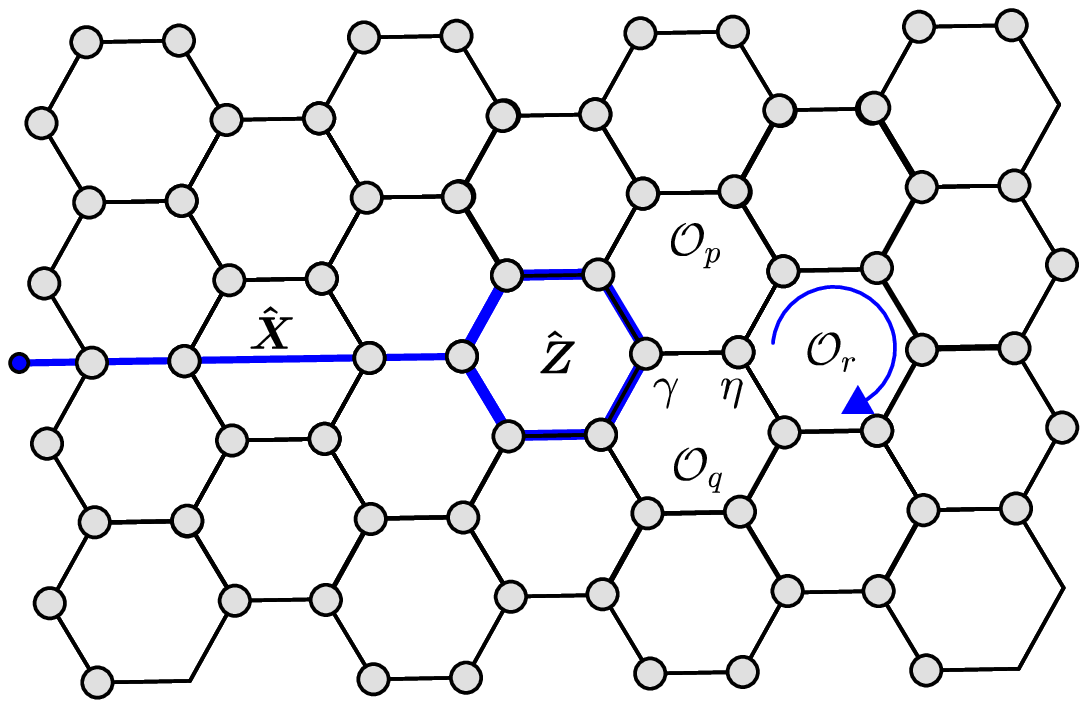}\\
\text{(a)}\\
\\
\includegraphics[trim = 7 200 160 40, clip = true, width=0.39\textwidth, angle = 0.]{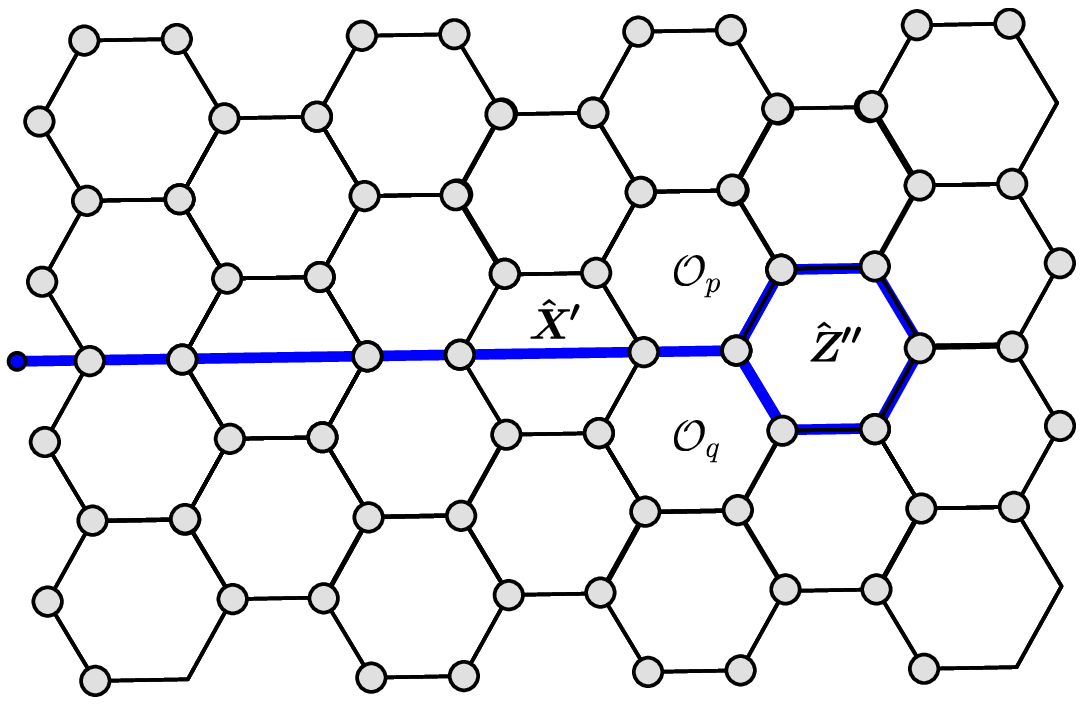}\\
\text{(b)}
\end{array}$
\caption{We may move a logical qubit defined by $\hat{X}$ and $\hat{Z}$ operators along a given sublattice. We first multiply the logical $\hat{Z}$ by $\hat{\mathcal{O}}_{r}$ and turn $\hat{\mathcal{O}}_{p}$ and $\hat{\mathcal{O}}_{q}$ into four-Majorana operators. After measuring $i\gamma\eta$ in the next code cycle, we extend the logical $\hat{X} \rightarrow \hat{X}\otimes i\gamma\eta$. Finally, we begin measuring $\hat{Z}$ in the next surface code cycle and restore $\hat{\mathcal{O}}_{p}$ and $\hat{\mathcal{O}}_{q}$ to six-Majorana operators.\label{fig:Moving_Qubits}}
\end{figure}

Consider the $A$-type logical qubit shown in Figure \ref{fig:Moving_Qubits}. To move the qubit one unit to the right, we perform the following sequence of measurements. We begin by multiplying the $\hat{Z}$ logical operator by the eigenvalue of the adjacent $r$ plaquette operator to perform the transformation:
\begin{align}
\hat{Z}\rightarrow\hat{Z}'\equiv\hat{Z}\otimes\hat{\mathcal{O}}_{r}.
\end{align}
As the $r$ plaquette is being continuously measured, its eigenvalue $\mathcal{O}_{r} = \pm 1$ is known from the previous surface code cycle. In the next cycle, we stop measuring $\hat{\mathcal{O}_{r}}$
and measure the Majorana bilinear $i\gamma\eta$.  We then multiply the $\hat{X}$ operator by the measurement outcome, affecting the transformation 
\begin{align}
\hat{X}\rightarrow\hat{X}'\equiv\hat{X}\otimes i\gamma\eta.
\end{align}
In the final surface code cycle, we begin measuring the original $\hat{Z}$ stabilizer and continue to include the measurement of the $\hat{Z}$ stabilizer in all subsequent surface code cycles. Furthermore, we redefine the logical operator $\hat{Z}'$ as
\begin{align}
\hat{Z}'\rightarrow\hat{Z}'' \equiv \hat{Z}'\otimes\hat{Z}.
\end{align}
The initial qubit configuration and final outcome are depicted schematically in Figure \ref{fig:Moving_Qubits}.  This sequence of measurements has shifted the $A$-type qubit by moving its hexagonal end one unit to the right, and may generally be used to move an $A$- or $B$-type logical qubit within the $A$ or $B$ plaquettes, respectively.  

\begin{figure}
$\begin{array}{c}
\includegraphics[trim = 95 115 190 30, clip = true, width=0.3\textwidth, angle = 0.]{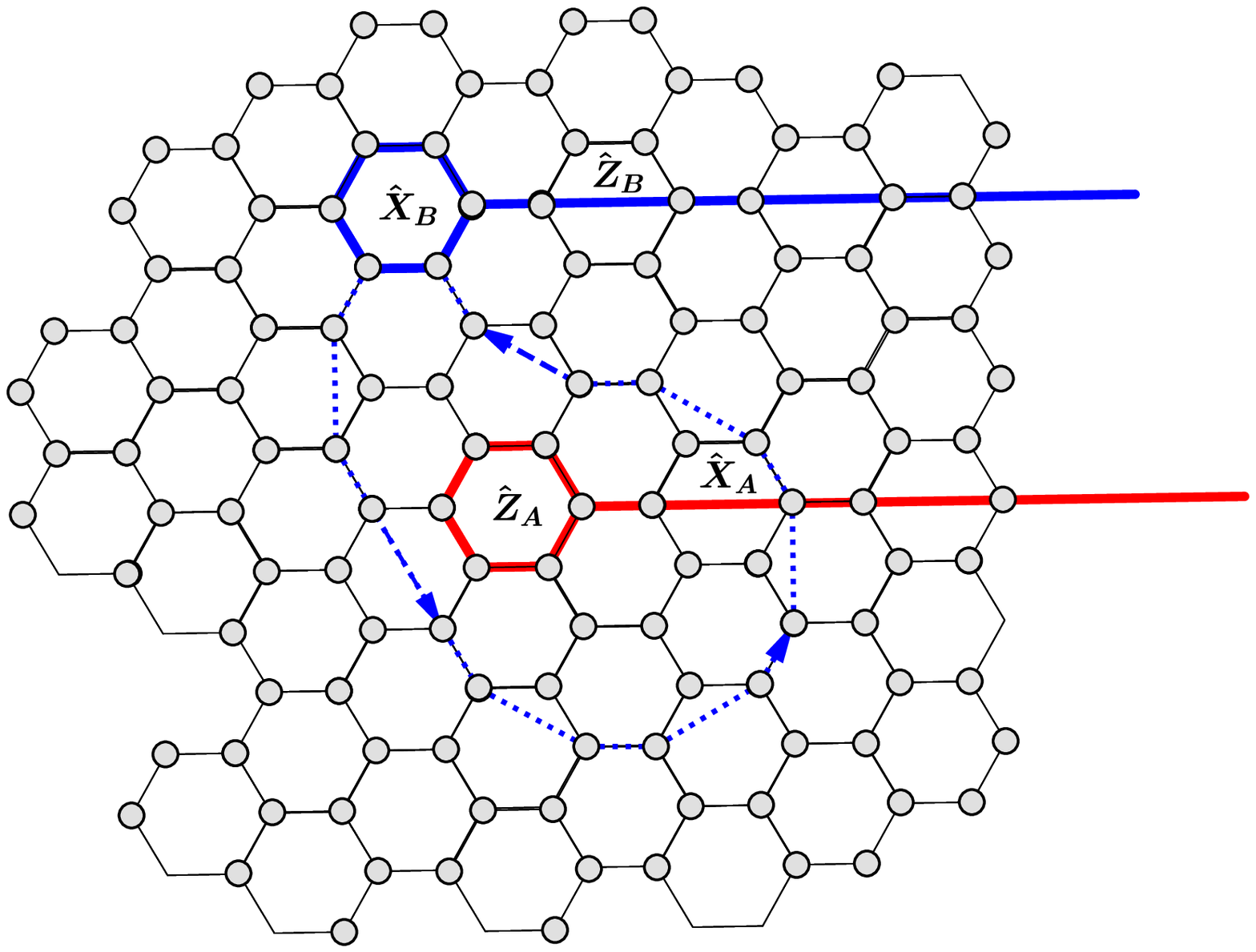}\\
\text{(a)}\\
\\
\,\includegraphics[trim = 90 55 90 20, clip = true, width=0.3\textwidth, angle = 0.]{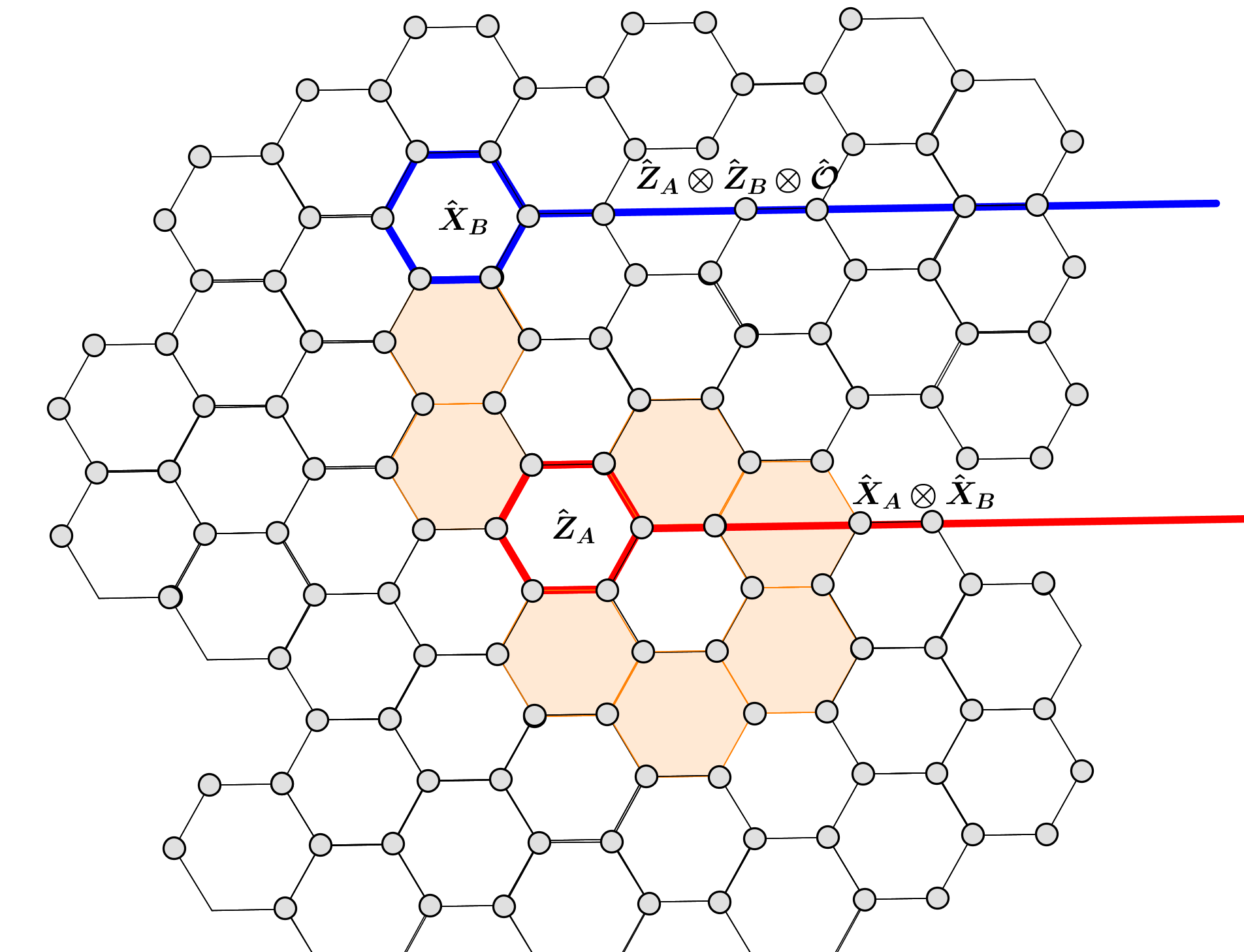}\\
\text{(b)}
\end{array}$
\caption{\textbf{CNOT Gate}. Braiding two logical qubits to perform a logical CNOT.  In (a), a possible trajectory for braiding the first qubit around the second is indicated by the dotted line. Since the two qubits live on distinct sublattices, the braiding procedure induces the transformation $\hat{X}_{A}\rightarrow\hat{X}_{A}\otimes\hat{X}_{B}$ and $\hat{Z}_{B}\rightarrow\hat{Z}_{A}\otimes\hat{Z}_{B}\otimes\hat{\mathcal{O}}$, where $\hat{\mathcal{O}}$ is the product of the colored plaquettes shown. This performs a CNOT transformation on the braided qubit.\label{fig:CNOT_Gate}}
\end{figure}

We may now braid pairs of logical qubits to perform a CNOT gate in the Majorana surface code.  The simplest CNOT  that we may realize is between two distinct types of qubits, taking the $A$ qubit as the control, as shown in Figure \ref{fig:CNOT_Gate}.  Since the qubits are distinct, braiding the $B$-type qubit -- with logical operators $\hat{X}_{B}$ and $\hat{Z}_{B}$ -- along a closed path $\ell$ enclosing the second qubit (i) multiplies the Wilson line of the $B$-type qubit by the anyon charge enclosed by $\ell$ and (ii)  multiplies the Wilson line of the $A$ qubit by the anyon charge of the $B$ qubit. This results in the transformation:
\begin{align}
\hat{X}_{A}\rightarrow \hat{X}_{A}\otimes\hat{X}_{B} \hspace{.25in} \hat{Z}_{B}\rightarrow \hat{Z}_{A}\otimes\hat{Z}_{B}\otimes\prod_{p\in\ell}\hat{\mathcal{O}}_{p}
\end{align}
where $\{\hat{\mathcal{O}}_{p}\}$ are $A$ and $C$-type plaquette operators enclosed by the braiding trajectory, as shown in Figure \ref{fig:CNOT_Gate}. Since the eigenvalues of the enclosed plaquette operators are known from the {previous} surface code cycle, we may implement the logical CNOT ($\hat{Z}_{A}\rightarrow \hat{Z}_{A}$, $\hat{Z}_{B}\rightarrow\hat{Z}_{A}\otimes\hat{Z}_{B}$) by multiplying the transformed $\hat{Z}_{B}$ by an appropriate sign.  In summary, the simplest braiding process between an $A$ and a $B$ logical qubit implements a CNOT on the $B$ qubit, with the $A$ qubit as the control. 

A CNOT between two logical qubits of the \emph{same} type may also be performed by appropriately braiding pairs of distinct types of logical qubits.  In this case, we will take one qubit as the control by convention and store the outcome of the CNOT gate in a third ancilla qubit. First, consider performing a CNOT gate on two $A$-type qubits.  To implement the CNOT, we prepare two additional ancilla qubits; the first is an $A$ qubit prepared in the state $\ket{\varphi} \equiv [\ket{+_{z}} + \ket{-_{z}}]/\sqrt{2}$, while the second is a $B$ qubit prepared in the state $\ket{+_{x}}$, with $\ket{\pm_{z}}$ and $\ket{\pm_{x}}$ the eigenstates of the logical $Z$ and $X$ operators, respectively. Both ancilla qubits are prepared by measuring a $+1$ eigenvalue for the Wilson line joining the pair of plaquettes of the appropriate qubit. For the $A$ ($B$) qubit, this projects onto an eigenstate of the logical $X$ ($Z$) operator, and produces the desired ancilla states.

\begin{figure}
$\begin{array}{c}
\includegraphics[trim = 0 115 200 10, clip = true, width=0.34\textwidth, angle = 0.]{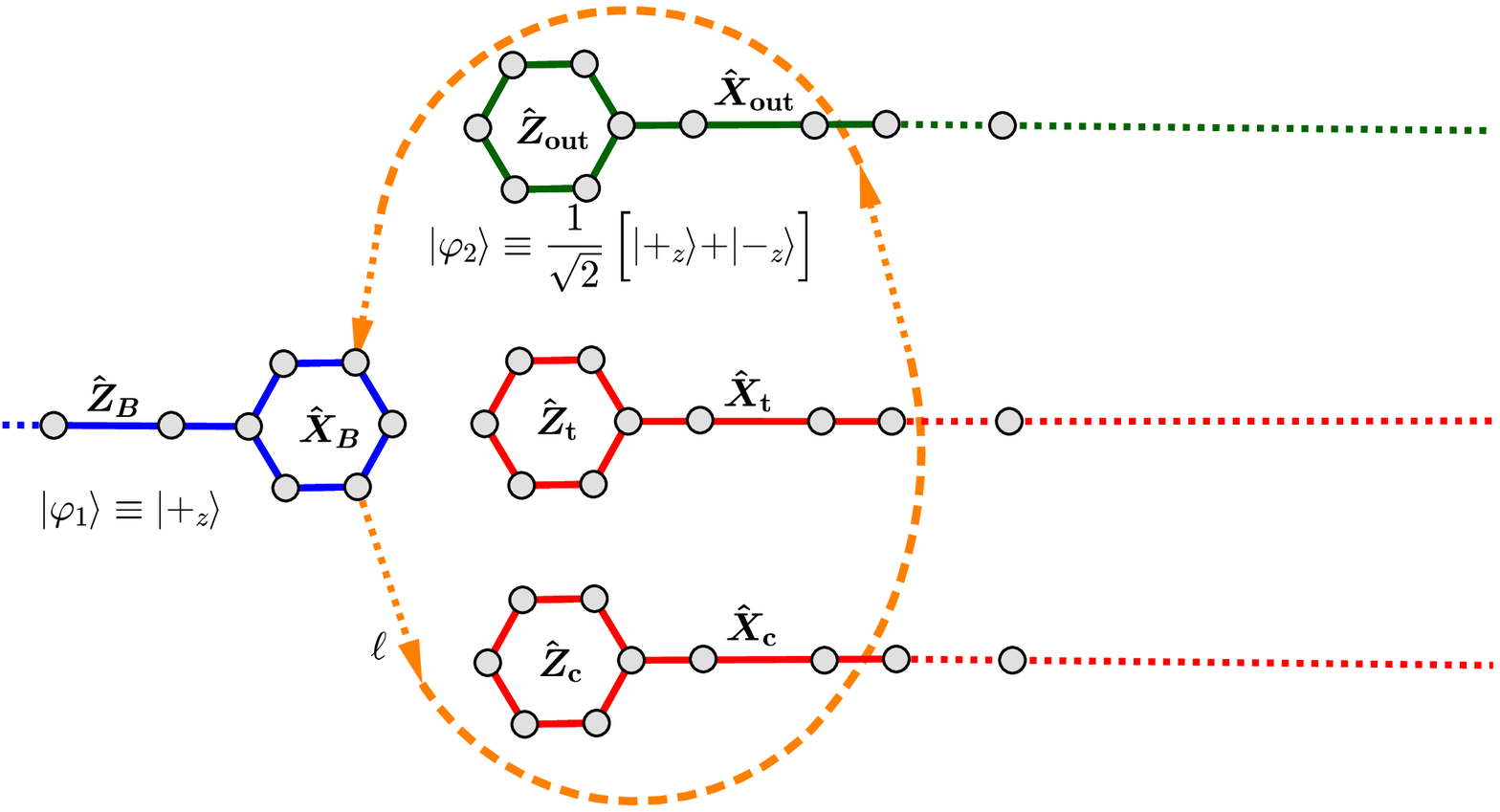}\\
\text{(a)}\\
\includegraphics[trim = 0 83 142 10, clip = true, width=0.36\textwidth, angle = 0.]{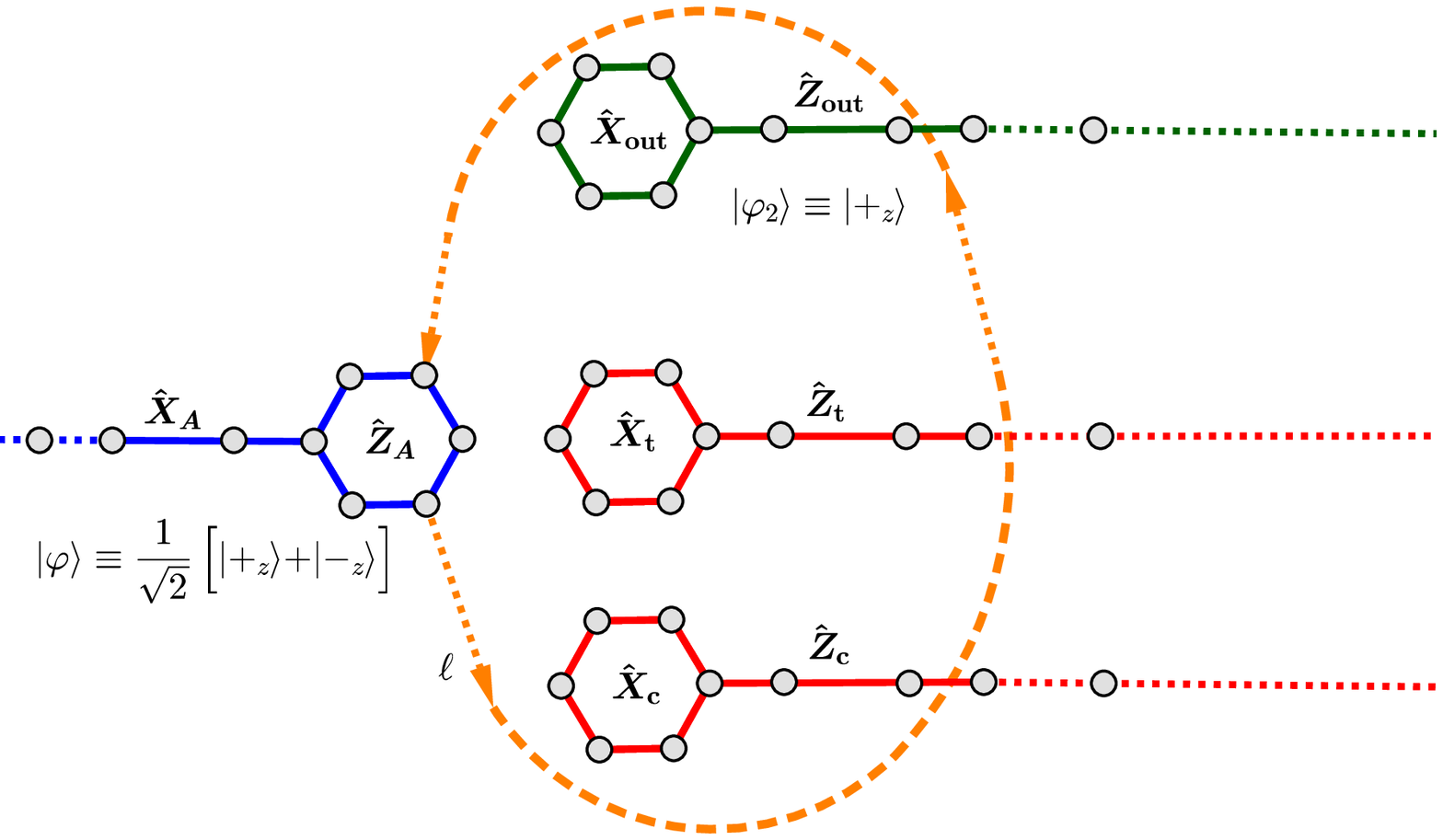}\\
\text{(b)}
\end{array}$
\caption{Braiding processes that implement the transformation (a) $\hat{Z}_{a}\rightarrow\hat{Z}_{a}\otimes\hat{Z}_{c}\otimes\hat{Z}_{t}\otimes\hat{Z}_{\mathrm{out}}$ up to an overall sign, as determined by the product of the remaining plaquette operators enclosed by the path $\ell$, and (b) $\hat{Z}_{\mathrm{out}}\rightarrow\hat{Z}_{\mathrm{out}}\otimes\hat{Z}_{A}$, $\hat{Z}_{t}\rightarrow\hat{Z}_{t}\otimes\hat{Z}_{A}$, $\hat{Z}_{c}\rightarrow\hat{Z}_{c}\otimes\hat{Z}_{A}$. The two braids are used to realize CNOT gates between two (a) $A$-type and (b) $B$-type logical qubits, respectively. By convention, we take the lowest qubit enclosed by the braiding trajectory to be the control for the logical CNOT.\label{fig:ZZ_CNOT_Gate}}
\end{figure}

We now represent a complete basis of the four-qubit states as $\ket{z_{B}, z_{c}, z_{t}, z_{\mathrm{out}}}$, referring to the eigenvalues of the logical $Z$ operators of the $B$ ancilla, the control, the target and the ancilla $A$ qubits, respectively. We start out with an initial state $\ket{\psi_{\mathrm{init}}}$ of the form:
\begin{align}
\ket{\psi_{\mathrm{init}}} \equiv \frac{1}{\sqrt{2}}\Big[\ket{+, z_{c}, z_{t}, +} + \ket{+, z_{c}, z_{t}, -}\Big]
\end{align}
Next, we braid the $B$ ancilla qubit around all three remaining qubits as shown in Figure \ref{fig:ZZ_CNOT_Gate}a. Up to an overall sign determined by the eigenvalues of plaquette operators enclosed by the braiding trajectory that are known from previous surface code cycles, this braiding implements the transformation $\hat{Z}_{B} \rightarrow \hat{Z}_{B}\otimes\hat{Z}_{c}\otimes\hat{Z}_{t}\otimes\hat{Z}_{\mathrm{out}}$ on the logical $Z$ of the ancilla qubit, where $\hat{Z}_{c}$, $\hat{Z}_{t}$ and $\hat{Z}_{\mathrm{out}}$ are the logical $Z$ operators for the control, target, and ancilla $A$-type qubits, respectively.  The final state we obtain is then of the form:
\begin{align}
\ket{\psi_{\mathrm{final}}} = \frac{1}{\sqrt{2}}\Big[\ket{z_{c}z_{t}, z_{c}, z_{t}, +} + \ket{-z_{c}z_{t}, z_{c}, z_{t}, -}\Big]
\end{align}
This braiding process is convenient, as a measurement of the state of the $B$ qubit can determine whether the the state of the $A$ \emph{ancilla} contains the correct outcome of the CNOT operation. If we now measure the logical $Z$ of the $B$ qubit and obtain $\hat{Z}_{B} = +1$ then we project onto a state with $z_{c}z_{t} = z_{\mathrm{out}}$. In this case, the $A$ ancilla qubit contains the correct outcome of the CNOT between the other $A$ qubits.  If $\hat{Z}_{B} = -1$, however, then $z_{c}z_{t} = -z_{\mathrm{out}}$ and the $A$ ancilla contains the opposite of the correct CNOT outcome. In this case, we may act with $\hat{X}_{\mathrm{out}}$ on the $A$ ancilla qubit in the surface code software \cite{Fowler_Martinis} to obtain the desired final state.

A similar process may be used to perform logical CNOT's between two $B$ qubits; now, we prepare an $A$ ancilla qubit and a $B$ ancilla qubit in the states shown in Figure \ref{fig:ZZ_CNOT_Gate}b.  After braiding the ancilla $A$ qubit around the control, target, and ancilla $B$ qubits, if we measure $\hat{X}_{A} = +1$, then the $B$ ancilla contains the desired outcome of the CNOT operation.  Again, by convention, we take the control qubit to be the first one enclosed by the braiding trajectory, as shown in Figure \ref{fig:ZZ_CNOT_Gate}b.

{\bf Hadamard gate:} The Hadamard is a single-qubit gate taking the matrix form:
\begin{align}
\hat{H} = \frac{1}{\sqrt{2}}\left(
\begin{array}{cc}
1 & 1\\
1 & -1
\end{array}
\right)
\end{align}
The action of a Hadamard is to exchange the logical $\hat{X}$ and $\hat{Z}$ operators so that $\hat{H}\hat{X}\hat{H}^{\dagger} = \hat{Z}$ and $\hat{H}\hat{Z}\hat{H}^{\dagger} = \hat{X}$. As the logical $\hat{X}$ and $\hat{Z}$ are defined oppositely on different types of qubits, a Hadamard operation in the bosonic surface code corresponds to an electric/magnetic duality transformation that interchanges star and plaquette operators in the toric code. In the ordinary surface code, such a transformation is quite difficult to implement, requiring a series of Hadamards on physical qubits enclosing the logical qubit so as to interchange the $\hat{X}$ and $\hat{Z}$ stabilizers, followed by physical swap gates in order to correctly patch the transformed logical qubit back into the remaining surface code array \cite{Fowler_Hadamard}.  As lattice symmetries permute the anyons in the Majorana plaquette model, however, the logical Hadamard may be realized in the Majorana surface code by simply moving a logical qubit between \emph{distinct} plaquettes.

\begin{figure}[t]
$\begin{array}{c}
\includegraphics[trim = 0 160 110 0, clip = true, width=0.35\textwidth, angle = 0.]{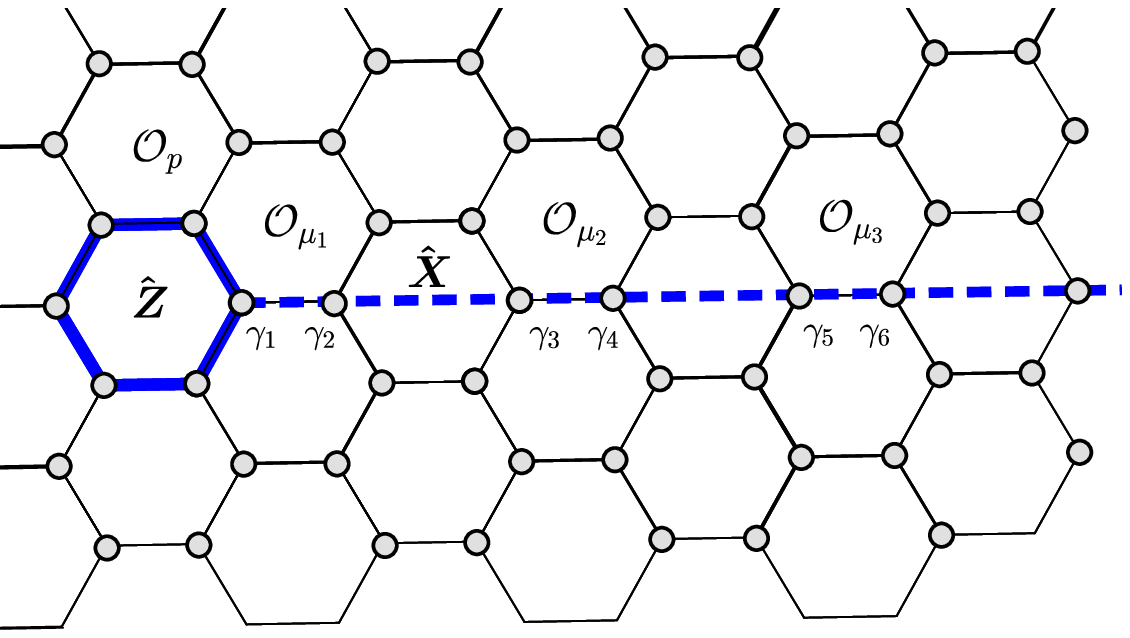} \\
\\
\text{(a)}\\
\\
\includegraphics[trim = 0 135 77 0, clip = true, width=0.35\textwidth, angle = 0.]{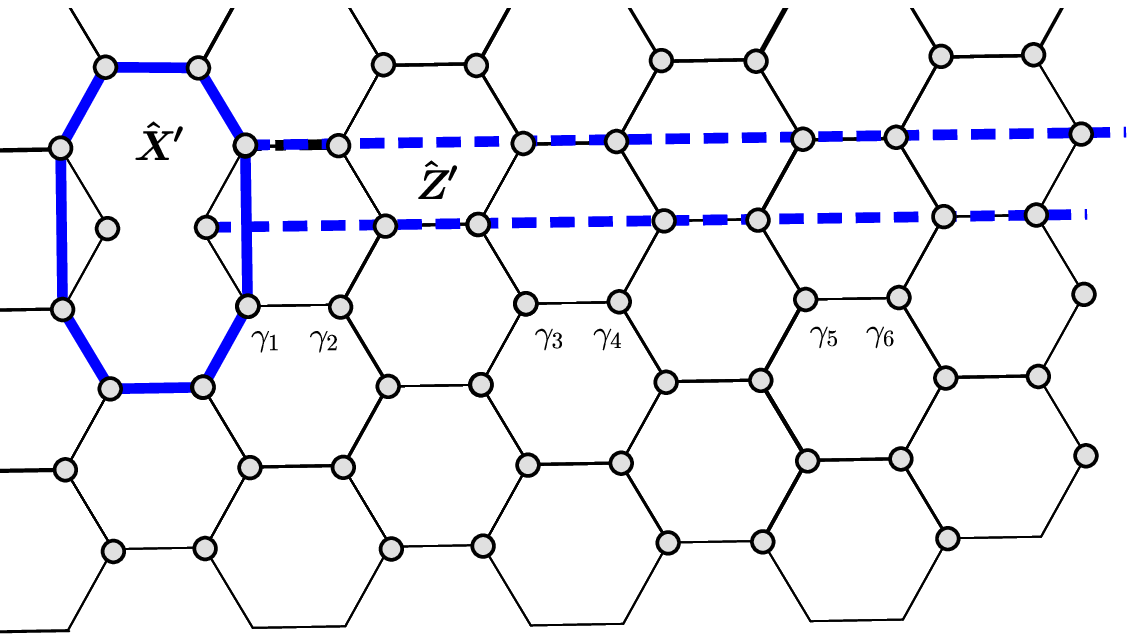} \\
\text{(b)}\\
\\
\includegraphics[trim = 0 195 77 0, clip = true, width=0.35\textwidth, angle = 0.]{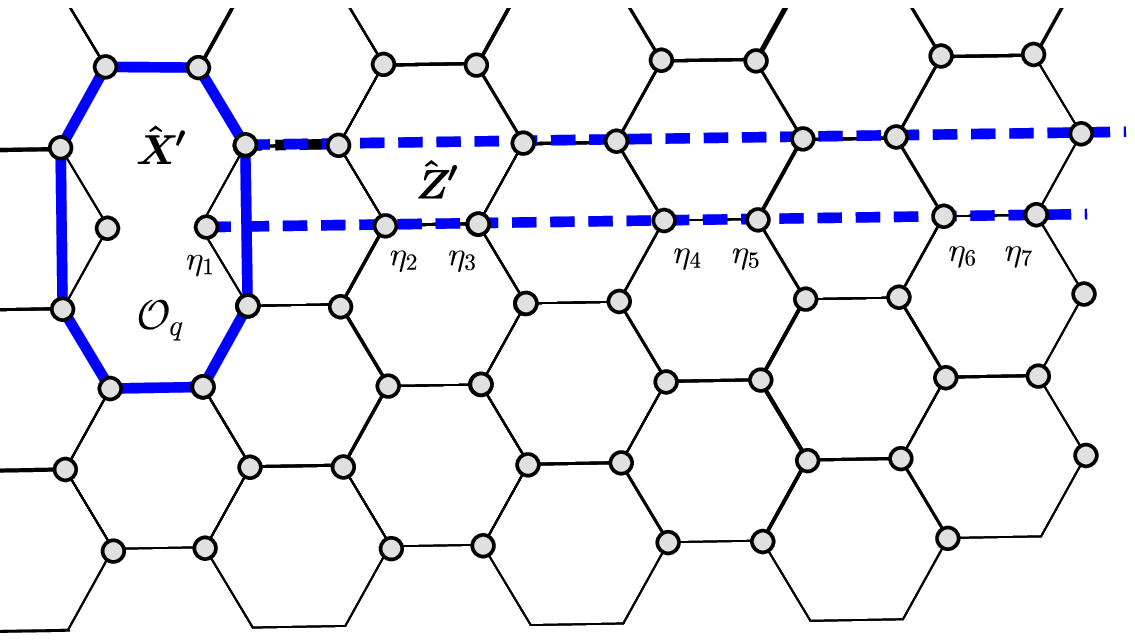} \\
\text{(c)}\\
\\
\includegraphics[trim = 0 245 77 0, clip = true, width=0.35\textwidth, angle = 0.]{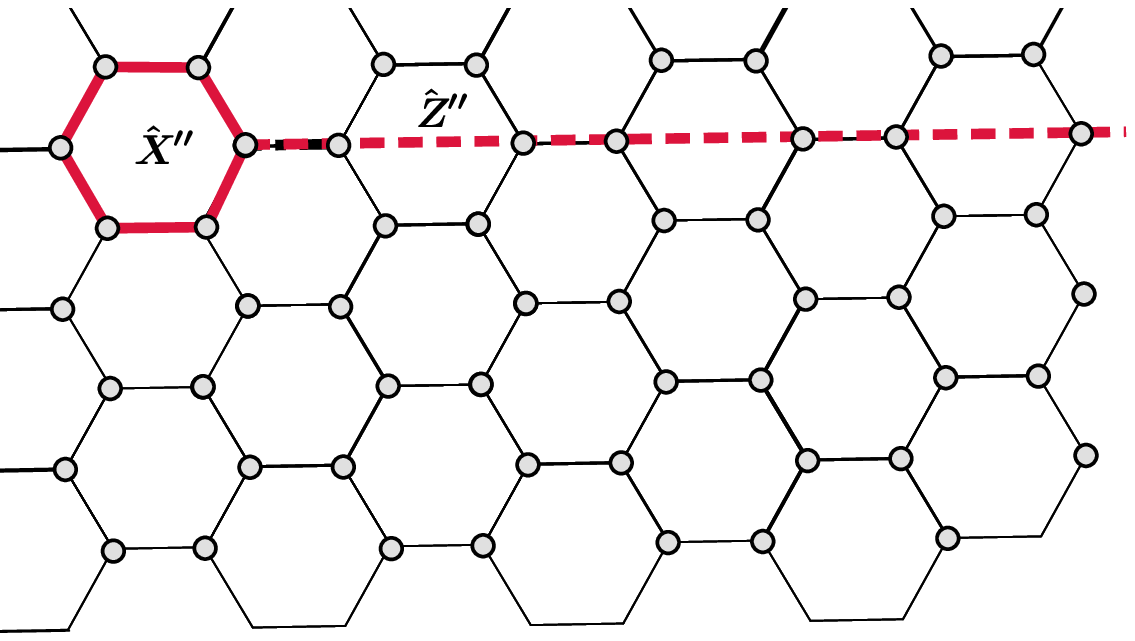} \\
\text{(d)}
\end{array}$
\caption{\textbf{Hadamard gate}. A logical Hadamard is performed by transferring a qubit between distinct sublattices, so that the logical $\hat{X}$ and $\hat{Z}$ operators are exchanged. We do this by taking the qubit in (a) and multiplying by logical $\hat{X}$ by the plaquette operators $\{\hat{\mathcal{O}}_{\mu_{k}}\}$ and the logical $\hat{Z}$ by $\hat{\mathcal{O}}_{p}$ and ceasing measurement of the fermion parity of plaquette $p$, yielding the operators shown in (b). Next, we measure the product $(i\eta_{1}\eta_{2})(i\eta_{3}\eta_{4})\cdots$ and and $\hat{\mathcal{O}}_{q}$ and multiply with $\hat{Z}'$ and $\hat{X}'$, respectively. The final result is shown in (d).\label{fig:Qubit_Transfer}}
\end{figure}

We implement the logical Hadamard by the procedure shown in Figure \ref{fig:Qubit_Transfer}.  Consider an $A$-type logical qubit. We multiply the logical $\hat{X}$ operator of the qubit, defined by the Wilson line in Figure \ref{fig:Qubit_Transfer}a, by the product of the adjacent plaquette operators $\{\hat{\mathcal{O}}_{\mu_{k}}\}$ extending between the hexagonal ends of the qubit. The eigenvalues of these plaquette operators are known from previous surface code cycles.  This operation implements the transformation:
\begin{align}
\hat{X}\rightarrow\hat{Z}' \equiv \hat{X}\otimes\prod_{k}\hat{\mathcal{O}}_{\mu_{k}}.
\end{align}
At the same time, we multiply the logical $\hat{Z}$ by the adjacent plaquette operator $\hat{\mathcal{O}}_{p}$, shown in Figure \ref{fig:Qubit_Transfer}a, that borders the logical qubit above:
\begin{align}
\hat{Z}\rightarrow\hat{X}'\equiv \hat{Z}\otimes\hat{\mathcal{O}}_{p}.
\end{align}
In subsequent surface code cycles, we stop measuring the eigenvalue of $\hat{\mathcal{O}}_{p}$. We implement a similar transformation on the other hexagonal end of the logical qubit, by stopping the measurement of the plaquette operator above the other qubit `hole'. The end result, after performing these operations, is shown in Figure \ref{fig:Qubit_Transfer}b. The solid and dashed blue lines indicate the products of the Majorana fermions on the appropriate sites that define the $\hat{X}'$ and $\hat{Z}'$ operators, respectively. 

In the next surface code cycle, we measure the product $(i\eta_{1}\eta_{2})(i\eta_{3}\eta_{4})\cdots$ of the Majorana fermions along the lower `string' that defines the $\hat{Z}'$ operator; this measurement commutes with $\hat{X}'$ since the two operators do not overlap, as shown in Figure \ref{fig:Qubit_Transfer}c. Afterwards, we measure $\hat{\mathcal{O}}_{q}$, as well as the plaquette operator $\hat{\mathcal{O}}_{h}$ for the other `hole' of the original logical qubit. Then, we may perform the following transformations on the logical $\hat{X}'$ and $\hat{Z}'$ operators:
\begin{align}
&\hat{X}'\rightarrow\hat{X}'' \equiv \hat{X}'\otimes\hat{\mathcal{O}}_{q}\\
&\hat{Z}'\rightarrow\hat{Z}'' \equiv \hat{Z}'\otimes\prod_{\ell}(i\eta_{2\ell-1}\eta_{2\ell}).
\end{align}
This yields the logical qubit shown in Figure \ref{fig:Qubit_Transfer}d. In subsequent surface code cycles, we continue measuring the eigenvalues of $\hat{\mathcal{O}}_q$ and $\hat{\mathcal{O}}_{h}$.  Since the logical $\hat{Z}$ and $\hat{X}$ operators are defined differently on the $A$ and $B$-type plaquettes, our procedure for transforming our $A$ qubit into a $B$ qubit implements a logical Hadamard gate.  An identical protocol may be used to perform a Hadamard on a logical $B$ qubit.

{\bf $\boldsymbol{S}$ and $\boldsymbol{T}$-gates:}
Finally, we implement the logical $S$- and $T$-gates, described by the following single-qubit operations:
\begin{align}
\hat{S} = \left(
\begin{array}{cc}
1 & 0\\
0 & i
\end{array}
\right) 
\hspace{.25in}
\hat{T} = \left(
\begin{array}{cc}
1 & 0\\
0 & e^{i\pi/4}
\end{array}
\right) 
\end{align}
As demonstrated in \cite{Fowler_Martinis}, it is possible to realize these gates by performing a series of logical Hadamard and CNOT gates between the logical qubit and an appropriate logical ancilla qubit. Here, we first discuss the $S$- and $T$-gate implementations, given the appropriate ancilla qubit, before outline a procedure for creating these logical ancillas in the surface code.

To implement an $S$-gate, we prepare a logical ancilla in the state
\begin{align}
\ket{\varphi_{S}} \equiv \frac{1}{\sqrt{2}}\Big[\ket{+_{z}} + i\ket{-_{z}}\Big]
\end{align}
Then, if $\ket{\Psi}$ is the state of the logical qubit of interest, the following sequence of logical Hadamard and CNOT gates implements the transformation $\ket{\Psi}\rightarrow\hat{S}\ket{\Psi}$ \cite{Fowler_Martinis}:
\begin{figure}[h]
\includegraphics[trim = 0 195 145 0, clip = true, width=0.34\textwidth, angle = 0.]{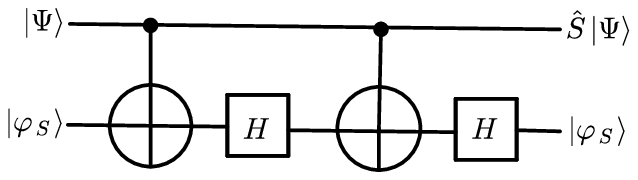}
\end{figure}

To perform a $T$-gate, we first prepare a logical ancilla in the state
\begin{align}
\ket{\varphi_{T}} \equiv \frac{1}{\sqrt{2}}\Big[\ket{+_{z}} + e^{i\pi/4}\ket{-_{z}}\Big].
\end{align}
The $T$-gate is then implemented via a probabilistic circuit. We perform a CNOT between the ancilla and the logical qubit of interest, and then measure the logical $\hat{Z}$ of the qubit.  Depending on the measurement outcome, we implement an $S$-gate as shown below:
\begin{figure}[h]
\includegraphics[trim = 0 195 145 0, clip = true, width=0.34\textwidth, angle = 0.]{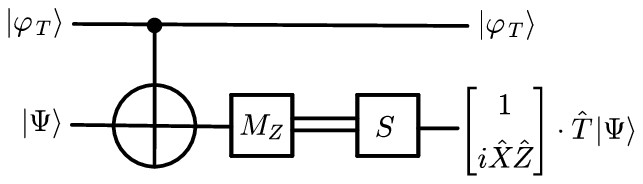}
\end{figure}
If the measurement outcome $M_{Z} = +1$, then we obtain the correct output $\hat{T}\ket{\Psi}$; otherwise, if $M_{Z} = -1$, then we have performed the transformation $\ket{\Psi}\rightarrow \hat{X}\hat{T}^{\dagger}\ket{\Psi}$. In this case, we implement an $S$-gate on the logical qubit and obtain the final state $i\hat{X}\hat{Z}\hat{T}\ket{\Psi}$. The action of the operator $i\hat{X}\hat{Z}$ may be undone in the surface code software to implement a pure $T$-gate on the logical qubit \cite{Fowler_Martinis}.

\begin{figure}[t]
\includegraphics[trim = 25 142 140 0, clip = true, width=0.25\textwidth, angle = 0.]{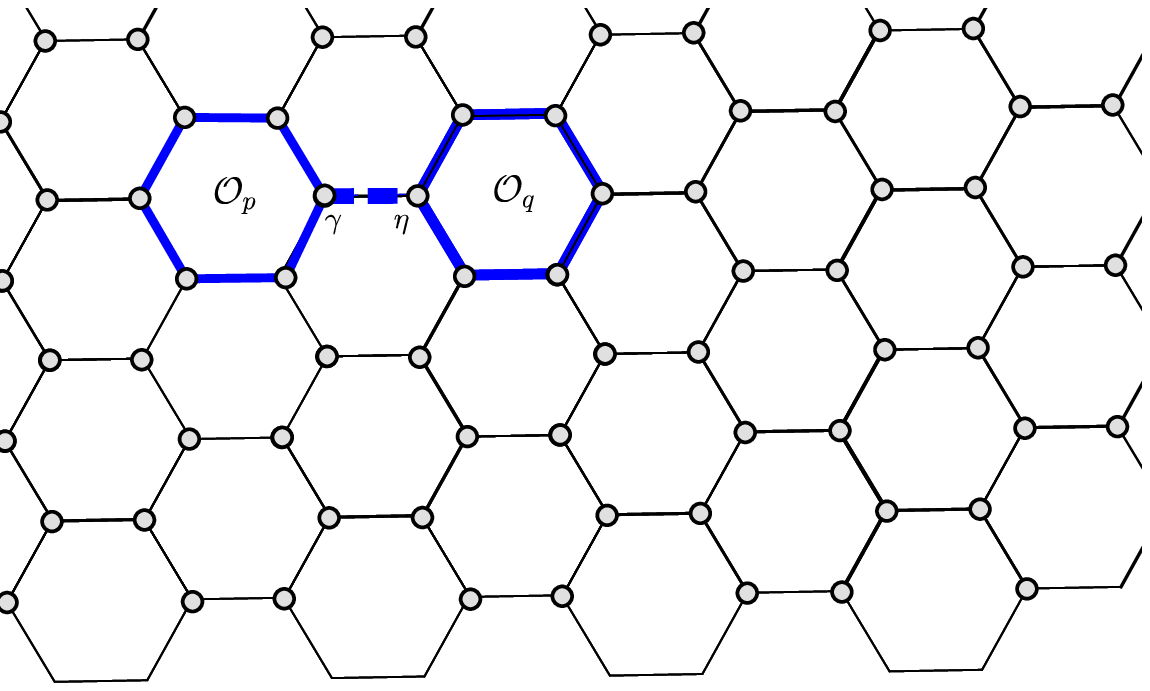}
\caption{\textbf{$\boldsymbol{S}$- and $\boldsymbol{T}$-gate Ancilla Preparation}. We create the $\ket{\varphi_{S}}$ and $\ket{\varphi_{T}}$ ancilla states, needed to realize logical $S$- and $T$-gates by preparing the ``short qubit" \cite{Fowler_Martinis} shown above. We cease stabilizer measurements on two adjacent plaquettes $p$ and $q$. In the next surface code cycle, we perform a rotation of the two-level system defined by $i\gamma\eta$. Finally, we enlargen the logical qubit by extending one end of the qubit, to guarantee stability against noise.\label{fig:Ancilla}}
\end{figure}

To realize the above implementations, we may prepare logical ancilla qubits in the states $\ket{\varphi_{T}}$ and $\ket{\varphi_{S}}$ as follows.  First, we create a ``short qubit" \cite{Fowler_Martinis} by ceasing the fermion parity measurement on two adjacent plaquettes $p$,  $q$ belonging to the same sublattice, as shown in Figure \ref{fig:Ancilla}a. For this qubit, let $\hat{X} \equiv \hat{\mathcal{O}}_{p}$ and $\hat{Z} \equiv i\gamma\eta$.  The qubit is initialized to a state $\ket{\Psi_{\pm}}$ such that $\hat{X}\ket{\Psi_{\pm}} = \pm\ket{\Psi}$.  In a basis of eigenstates of the logical $\hat{Z}$, the qubit state takes the form $\ket{\Psi_{\pm}} = (\ket{+_{z}} \pm \ket{-_{z}})/\sqrt{2}$. Now, we assume that the two-level system formed by the pair of Majorana fermions $\gamma$ and $\eta$ can be manipulated by performing a rotation 
\begin{align}
\hat{R}(\theta) = \left(
\begin{array}{cc}
1 & 0\\
0 & e^{i\theta}
\end{array}
\right)
\end{align} 
that acts in the basis of $\ket{\pm_{z}}$ states. This may be implemented by using the phase of the adjacent superconducting islands to tune the coupling between the Majorana zero modes \cite{Jiang}.
To prepare the state $\ket{\varphi_{S}}$, we perform the rotation $\hat{R}\left(\pm{\pi}/{2}\right)\ket{\Psi_{\pm}}$ in the next surface code cycle, while to prepare $\ket{\varphi_{T}}$, we perform the operation $\hat{R}\left((2\pi\pm\pi)/4\right)\ket{\Psi_{\pm}}$. Afterwards, to guarantee the stability of the qubit against noise generated by the environment, we increase the length of the logical $\hat{X}$ operator by extending one end of the logical qubit, as discussed in detail previously and shown schematically in Figure \ref{fig:Moving_Qubits}.  In practice, a high-fidelity implementation of the $S$- and $T$-gates requires that the ``short qubits" are put through a distillation circuit, as discussed in \cite{Fowler_Martinis}, which may be implemented using a sequence of logical CNOT gates with other ancilla logical qubits.\\

We have presented a two-dimensional model of interacting Majorana fermions that realizes a novel type of $Z_{2}$ topological order with a microscopic $S_{3}$ anyon symmetry.  The required multi-fermion interactions in the plaquette model are naturally generated by phase-slips in an array of phase-locked $s$-wave superconducting islands on a TI surface. Based on this physical realization, we propose the Majorana surface code and provide the necessary measurement protocols and gate implementations for universal quantum computation. The Majorana surface code provides substantially increased error tolerance, reduced overhead, and simpler logical gate implementations over a surface code with bosonic physical qubits.  We are optimistic that the Majorana fermion surface code will be physically implemented, and may provide an advantageous platform for fault-tolerant quantum computation.  \\

\begin{acknowledgments}
We thank Patrick Lee for helpful comments and discussion.  This work was supported by the Packard Foundation (LF), and the DOE Office of Basic Energy Sciences, Division of Materials Sciences and Engineering under Award No. DE-SC0010526 (SV and TH).
\end{acknowledgments}

SV and LF conceived and developed the Majorana fermion surface code for universal quantum computation and wrote the manuscript. TH contributed to the analysis of the Majorana plaquette model.

\section{Supplemental Material}
\section{Braiding Statistics of Excitations in the Majorana Plaquette Model}
\begin{figure}
\includegraphics[trim = 0 47 92 0, clip = true, width=0.34\textwidth, angle = 0.]{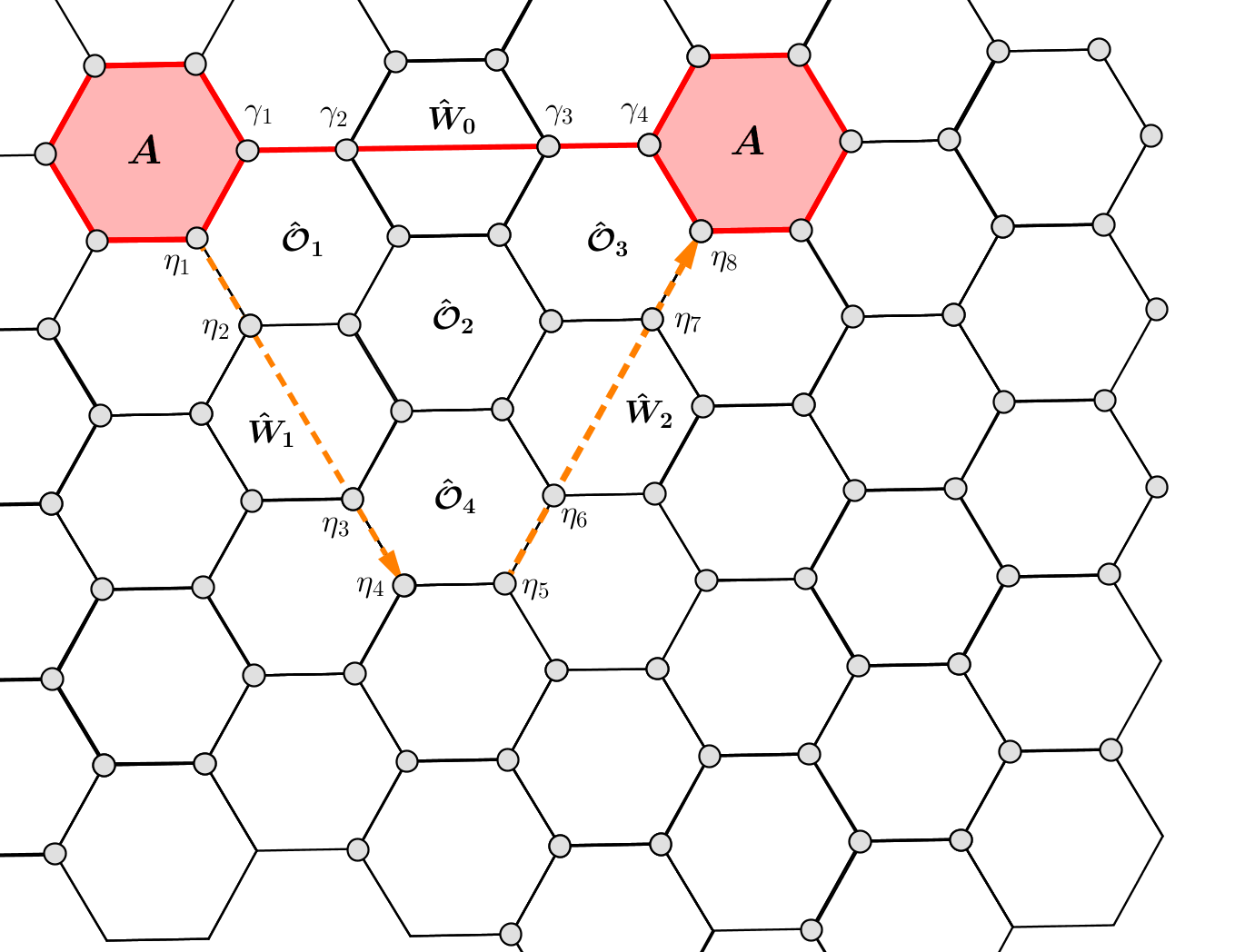}
\caption{The bosonic self-statistics of the $A$, $B$, and $C$ excitations in the Majorana plaquette model may be determined from the algebra of the Wilson lines used to create and move excitations.}\label{fig:Self_Statistics_Supp_Mat}
\end{figure}

We may braid the excitations of the Majorana plaquette model to determine their statistics.  First, consider acting on the ground-state $\ket{\Psi}$ of the plaquette model with a Wilson line $\hat{W}_{0} = (i\gamma_{1}\gamma_{2})(i\gamma_{3}\gamma_{4})$ to create a pair of excitations on the $A$-type plaquettes, as shown in Figure \ref{fig:Self_Statistics_Supp_Mat}.  Let $\ket{\Phi} \equiv \hat{W}_{0}\ket{\Psi}$.  We braid the two excitations by performing the operation $\hat{W}_{2}\hat{W}_{0}\hat{W}_{1}\ket{\Phi}$ where
\begin{align}
&\hat{W}_{1} \equiv (i\eta_{1}\eta_{2})(i\eta_{3}\eta_{4})\\
&\hat{W}_{2} \equiv (i\eta_{5}\eta_{6})(i\eta_{7}\eta_{8})
\end{align}
However, we note that the exchange transformation may be rewritten as the product of the plaquette operators enclosed by the exchanging trajectory so that $\hat{W}_{2}\hat{W}_{0}\hat{W}_{1} = \hat{\mathcal{O}}_{1}\hat{\mathcal{O}}_{2}\hat{\mathcal{O}}_{3}\hat{\mathcal{O}}_{4}$.  Since $\hat{\mathcal{O}}_{i}\ket{\Phi} = \ket{\Phi}$ for the enclosed plaquettes, we conclude that:
\begin{align}
\hat{W}_{2}\hat{W}_{0}\hat{W}_{1}\ket{\Phi} = \ket{\Phi}
\end{align}
so that the $A$ excitations have bosonic self-statistics.  By symmetry, the remaining $B$ and $C$-type fundamental excitations also possess bosonic self-statistics.

\begin{figure}
\includegraphics[trim = 10 40 40 47, clip = true, width=0.4\textwidth, angle = 0.]{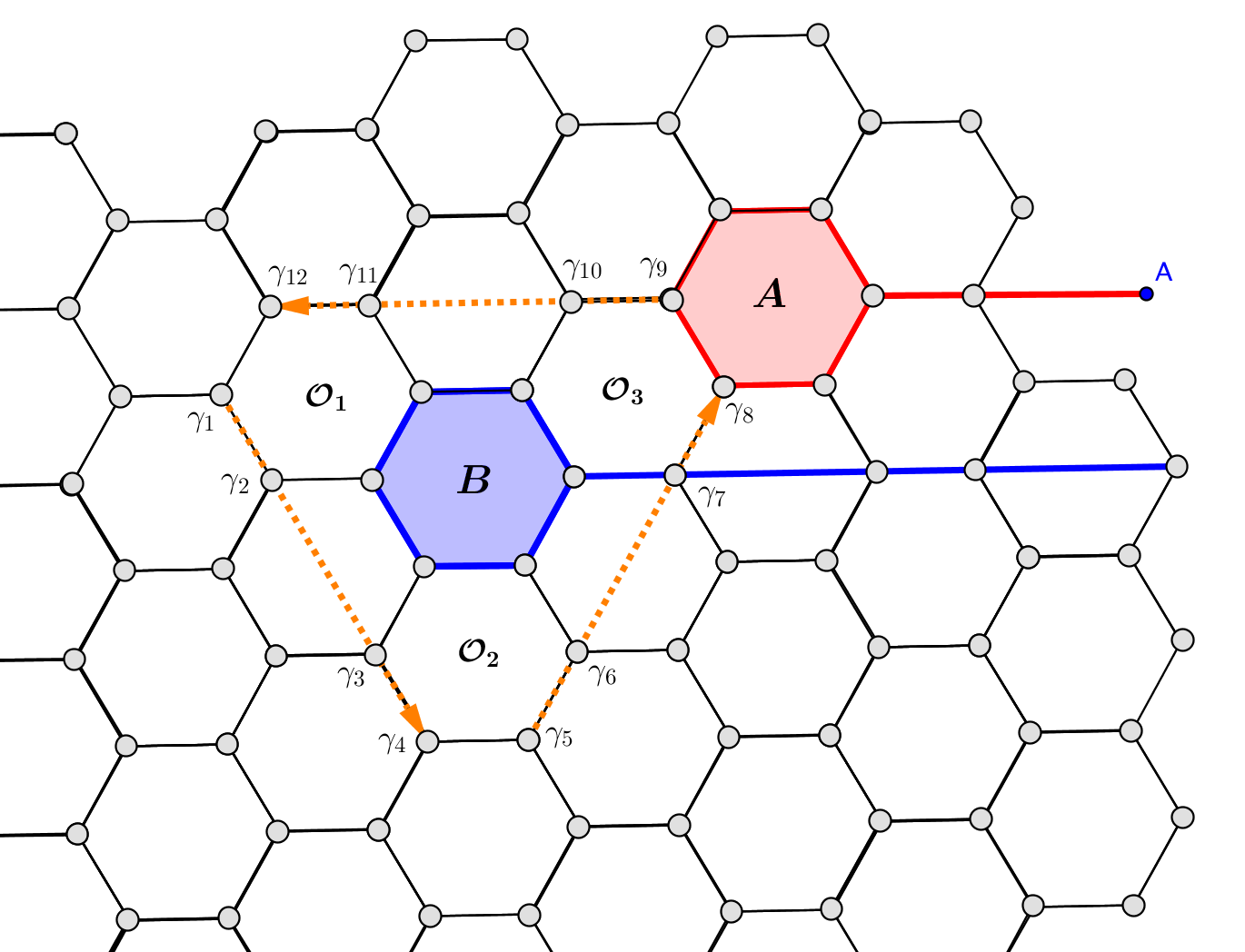}
\caption{Braiding two distinct fundamental excitations of the Majorana plaquette model. We braid the $A$ excitation around the $B$ excitation by acting with the product of the Majorana operators shown. Since this is equivalent to acting with the product of the plaquette operators enclosed by the trajectory $\mathcal{O}_{1}\mathcal{O}_{2}\mathcal{O}_{3}\mathcal{O}_{B}$, the braiding process results in an overall sign due to the enclosed B excitation.
}\label{fig:Braiding}
\end{figure}

We may determine the mutual statistics of the fundamental excitations by braiding distinct excitations, as shown in Figure \ref{fig:Braiding}.  In this case, the braiding procedure will yield an irreducible $\pi$ Berry phase due to the charge of the enclosed excitation.  Let $\ket{\tilde{\Phi}}$ be the particular state shown in Figure \ref{fig:Braiding}, with a pair of $A$ and $B$ excitations. We braid the $A$ excitation by acting with 
\begin{align}
\hat{W} \equiv \prod_{n = 1}^{6}(i\gamma_{2n-1}\gamma_{2n}) 
\end{align}
However, since this Wilson loop may also be written as the product of the enclosed plaquette operators$\hat{W} = \hat{\mathcal{O}}_{1}\hat{\mathcal{O}}_{2}\hat{\mathcal{O}}_{3}\hat{\mathcal{O}}_{B}$, we find that:
\begin{align}
\hat{W}\ket{\tilde{\Phi}} = \hat{\mathcal{O}}_{1}\hat{\mathcal{O}}_{2}\hat{\mathcal{O}}_{3}\hat{\mathcal{O}}_{B}\ket{\tilde{\Phi}} = -\ket{\tilde{\Phi}}
\end{align}
since $\hat{\mathcal{O}}_{j}\ket{\tilde{\Phi}} = \ket{\tilde{\Phi}}$ for $j = 1$, $2$, $3$ and $\hat{\mathcal{O}}_{B}\ket{\tilde{\Phi}} = -\ket{\tilde{\Phi}}$ is the enclosed $B$ anyon excitation. We conclude that each of the fundamental excitations are bosons with mutual semion statistics.  

\section{Projective Realization of the $S_{3}$ Anyon Symmetry}
In our Majorana plaquette model, the three fundamental excitations $A$, $B$ and $C$ live on distinct plaquettes of the honecyomb lattice. As a consequence, crystal symmetries of the honeycomb lattice permute the three types of fundamental excitations.  For example, we may take the generators of the $S_{3}$ permutation symmetry between the three anyons to be a $\pi/3$ rotation about the center of an A  ($R_{A}$) or a B plaquette ($R_{B}$); these generators implement the transformations $R_{A}: (A, B, C) \rightarrow (A, C, B)$ and $R_{B}: (A, B, C) \rightarrow (B, A, C)$, respectively.

The $S_{3}$ anyon symmetry may also be \emph{projectively} realized if the coupling $u < 0$ in the Majorana plaquette Hamiltonian (\ref{eq:Hamiltonian}), so that the $A$, $B$ and $C$ excitations occupy all plaquettes in the ground-state $\ket{\tilde{\Psi}_{0}}$. Excitations above the ground-state now correspond to the absence of a anyons occupying pairs of plaquettes of the same time.  However, transporting such an excitation around an elementary loop now encloses $\pi$-flux and multiples the final state by an overall sign. This process may be performed entirely with the $S_{3}$ generators, so that $(R_{A})^{6}\ket{\tilde{\Psi}_{0}} = (R_{B})^{6}\ket{\tilde{\Psi}_{0}} = -\ket{\tilde{\Psi}_{0}}$; therefore, the anyon symmetry is projectively realized in the `$\pi$-flux' regime of the Majorana plaquette model. The phase transition between the $u > 0$ and $u < 0$ regimes realizes dimension three `symmetry-breaking' \cite{Kitaev_2}.


\begin{thebibliography}{1}

\bibitem{Majorana}
E. Majorana, Nuovo Cimento, {\bf 14}, 171 (1937).

\bibitem{Alicea}
J. Alicea, Rep. Prog. Phys. {\bf 75}, 076501 (2012).

\bibitem{Beenakker_review}
C. W. J. Beenakker, Annu. Rev. Con. Mat. Phys. {\bf 4}, 113 (2013).

\bibitem{Fu_Kane}
L. Fu and C. L. Kane, Phys. Rev. Lett. {\bf 100}, 096407 (2008).

\bibitem{sarma}
R. M. Lutchyn, J. D. Sau, and S. Das Sarma, Phys. Rev. Lett. {\bf 105}, 077001 (2010)

\bibitem{oreg}
Yuval Oreg, Gil Refael, and Felix von Oppen, Phys. Rev. Lett. {\bf 105}, 177002 (2010)

\bibitem{palee}
A. C. Potter and P. A. Lee, Phys. Rev. Lett. {\bf 105}, 227003 (2010). 

\bibitem{chain}
S. Nadj-Perge, I. K. Drozdov, B. A. Bernevig, Ali Yazdani, Phys. Rev. B {\bf 88}, 020407(R) (2013). 

\bibitem{kouwenhoven}
V. Mourik, K. Zuo, S. M. Frolov, S. R. Plissard, E. P. A. M. Bakkers, and L. P. Kouwenhoven, 
Science {\bf 336}, 1003 (2012).

\bibitem{heiblum}
A. Das, Y. Ronen, Y. Most, Y. Oreg, M. Heiblum, H. Shtrikman, 
Nat. Phys. {\bf 8}, 887 (2012).

\bibitem{yazdani}
Stevan Nadj-Perge, Ilya K. Drozdov, Jian Li, Hua Chen, Sangjun Jeon, Jungpil Seo, Allan H. MacDonald, B. Andrei Bernevig, Ali Yazdani, 
Science {\bf 346}, 6209 (2014).

\bibitem{jia1}
Jin-Peng Xu, Mei-Xiao Wang, Zhi Long Liu, Jian-Feng Ge, Xiaojun Yang, Canhua Liu, Zhu An Xu, Dandan Guan, Chun Lei Gao, Dong Qian, Ying Liu, Qiang-Hua Wang, Fu-Chun Zhang, Qi-Kun Xue, and Jin-Feng Jia, Phys. Rev. Lett. {\bf 114}, 017001 (2015). 

\bibitem{law}
K. T. Law, P. A. Lee and T. K. Ng, 	Phys. Rev. Lett. {\bf 103}, 237001 (2009).

\bibitem{potter}
Jie Liu, Andrew C. Potter, K.T. Law, Patrick A. Lee, Phys. Rev. Lett. {\bf 109}, 267002 (2012). 

\bibitem{moore-read}
G. Moore and N. Read, Nucl. Phys. B {\bf 360}, 362 (1991).

\bibitem{read-green}
N. Read and D. Green, Phys. Rev. B {\bf 61}, 10267 (2000).

\bibitem{ivanov}
D. A. Ivanov, Phys. Rev. Lett. {\bf 86}, 268 (2001).


\bibitem{Toric_Code}
A. Kitaev, Ann. Phys. {\bf 303}, 2 (2003).

\bibitem{Nayak-rmp}
C. Nayak, S. H. Simon, A. Stern, M. Freedman, and S. Das Sarma, Rev. Mod. Phys. {\bf 80}, 1083 (2008).

\bibitem{Surf_Code_Kitaev}
S. Bravyi and A. Kitaev, Quantum Computers and Computing, {\bf 2}, 43 (2001).

\bibitem{Surf_Code_Freedman}
M. H. Freedman and D. A. Meyer, Found. Comput. Math. {\bf 1}, 325 (2001).

\bibitem{Raussendorf}
R. Raussendorf and J. Harrington, Phys. Rev. Lett. {\bf 98}, 190504 (2007).

\bibitem{Raussendorf_2}
R. Raussendorf, J. Harrington, and K. Goyal, New J. Phys. {\bf 9}, 199 (2007).

\bibitem{Fowler_Martinis}
A. G. Fowler, M. Mariantoni, J. M. Martinis and A. N. Cleland, Phys. Rev. A {\bf 86}, 032324 (2012).

\bibitem{Fowler_Surface_Code}
A. G. Fowler, A. M. Stephens, and P. Groszkowski, Phys. Rev. A {\bf 80}, 052312 (2009).

\bibitem{Error_Correction_Preskill}
E. Dennis, A. Kitaev, A. Landahl and J. Preskill, J. Math. Phys. {\bf 43}, 4452-4505 (2002).



\bibitem{Martinis} 
R. Barends, J. Kelly, A. Megrant, A. Veitia, D. Sank, E. Jeffrey, T. White, J. Mutus, A. Fowler, B. Campbell, \emph{et al.}, Nature {\bf 508}, 500 (2014).

\bibitem{Martinis_Surf_Code_Array}
J. Kelly, R. Barends, A. Fowler, A. Megrant, E. Jeffrey, T. White, D. Sank, J. Mutus, B. Campbell, Y. Chen, \emph{et al.}, Nature {\bf 519}, 66 (2015).

\bibitem{Steffen_Surf_Code_Array}
A. D. C\'{o}rcoles, E. Magesan, S. J. Srinivasan, A. W. Cross, M. Steffen, J. M. Gambetta and J. M. Chow, Preprint at http://arxiv.org/abs/1410.6419 (2014).


\bibitem{Kitaev}
A. Kitaev, Ann. Phys. {\bf 321}, 2 (2006).

\bibitem{Fu}
L. Fu, Phys. Rev. Lett. {\bf 104}, 056402 (2010). 

\bibitem{bravyi}
S. Bravyi, B. Terhal and B. Leemhuis, New. J. Phys. 12, 083039 (2010). 

\bibitem{XuFu}
C. Xu and L. Fu, Phys. Rev. B, {\bf 81}, 134435 (2010).

\bibitem{Teo}
J. C.Y. Teo, A. Roy and X. Chen, Phys. Rev. B {\bf 90}, 115118 (2014).

\bibitem{private}
Dale Van Harlingen, private communication. 


\bibitem{Top_Transmon}
F. Hassler, A. R. Akhmerov, and C. W. J. Beenakker, New Journal of Physics, {\bf 13}, 095004 (2011).

\bibitem{Schoelkopf}
L. Sun, L. DiCarlo, M. D. Reed, G. Catelani, Lev S. Bishop, D. I. Schuster, B. R. Johnson, G. A. Yang, L. Frunzio, L. Glazman, M. H. Devoret, and R. J. Schoelkopf, Phys. Rev. Lett. {\bf 108}, 230509, (2012).

\bibitem{Fowler_Hadamard}
A. G. Fowler, arXiv:quant-ph/1202.2369, (2012).

\bibitem{Yale_Group}
J. Koch, T. M. Yu, J. Gambetta, A. A. Houck, D. I. Schuster, J. Majer, A. Blais, M. H. Devoret, S. M. Girvin, and R. J. Schoelkopf, Phys. Rev. A {\bf 76}, 042319 (2007).

\bibitem{VanHeck1}
B. van Heck, F. Hassler, A. R. Akhmerov, and C. W. J. Beenakker, Phys. Rev. B 84, 180502(R) (2011) 

\bibitem{Hutzen}
R. Hutzen, A. Zazunov, B. Braunecker, A. L. Yeyati, and R. Egger, Phys. Rev. Lett. {\bf 109}, 166403  (2012). 

\bibitem{VanHeck2} 
B. van Heck, A. R. Akhmerov, F. Hassler, M. Burrello and C. W. J. Beenakker, New. J. Phys. {\bf 14} 035019 (2012). 

\bibitem{brinkman}
M. Veldhorst, M. Snelder, M. Hoek, T. Gang, V. K. Guduru, X. L.Wang, U. Zeitler, W. G. van der Wiel, A. A. Golubov, H. Hilgenkamp and A. Brinkman, 
Nature Mat. {\bf 11}, 417 (2012).

\bibitem{Future_Work}
S. Vijay and L. Fu, to be published. 

\bibitem{Jiang}
L. Jiang, C. L. Kane, and J. Preskill, Phys. Rev. Lett. {\bf 106}, 130504 (2011).

\bibitem{stanford} 
J.R. Williams, A. J. Bestwick, P. Gallagher, S. S. Hong, Y. Cui, A. S. Bleich, J. G. Analytis, I. R. Fisher, and D. Goldhaber-Gordon, 
Phys. Rev. Lett. {\bf 109}, 056803 (2012).

\bibitem{iop}
F. Yang, F. Qu, J. Shen, Y. Ding, J. Chen, Z. Ji, G. Liu, J. Fan, C. Yang, L. Fu, and L. Lu, 
Phys. Rev. B {\bf 86}, 134504 (2012).

\bibitem{molenkamp}
J. B. Oostinga, L. Maier, P. Schuffelgen, D. Knott, C. Ames, C. Brune, G. Tkachov, H. Buhmann, and L. W. Molenkamp, 
Phys. Rev. X {\bf 3}, 021007 (2013).

\bibitem{moler}
I. Sochnikov, A. J. Bestwick, J. R. Williams, T. M. Lippman, I. R. Fisher, D. Goldhaber-Gordon, J. R. Kirtley, and K. A. Moler, 
Nano Lett., {\bf 13}, 3086 (2013).

\bibitem{mason} 
S. Cho, B. Dellabetta, A. Yang, J. Schneeloch, Z. Xu, T. Valla, G. Gu, M.J. Gilbert, and N. Mason, Nature Communications, {\bf 4}, 1689 (2013)

\bibitem{harlingen}
C. Kurter, A. D. K. Finck, P. Ghaemi, Y. S. Hor, and D. J. Van Harlingen, Phys. Rev. B {\bf 90}, 014501 (2014).


\bibitem{SDS-PRL}
J. D. Sau, R. M. Lutchyn, S. Tewari and S. Das Sarma, Phys. Rev. Lett. {\bf 104}, 040502 (2010).

\bibitem{Alicea-PRB}
J. Alicea, Phys. Rev. B {\bf 81}, 125318 (2010). 

%


%

\bibitem{marcus}
A. P. Higginbotham, S. M. Albrecht, G. Kirsanskas, W. Chang, F. Kuemmeth, P. Krogstrup, T. S. Jespersen, J. Nygard, K. Flensberg, C. M. Marcus, 
arXiv:1501.05155


\bibitem{divincenzo}
B. M. Terhal, F. Hassler and D. P. DiVincenzo, Phys. Rev. Lett. {\bf 108}, 260504 (2012).

\bibitem{franz}
C. K. Chiu, D.I. Pikulin and M. Franz, arXiv:1411.5802

\bibitem{Minimum_Weight_Algorithm}
J. Edmonds, Can. J. Math. {\bf 17}, 449 (1965).

\bibitem{Fowler_Error_1}
D. S. Wang, A. G. Fowler, and L. C. L. Hollenberg, Phys. Rev. A {\bf 83}, 020302(R) (2011).

\bibitem{Fowler_Error_2}
D. S. Wang, A. G. Fowler, A. M. Stephens, and L. C. L. Hollenberg, Quant. Info. Comp. {\bf 10}, 456 (2010).

\bibitem{Fowler_Error_3}
A. G. Fowler, A. C. Whiteside, and L. C. L. Hollenberg, Phys. Rev. Lett. {\bf 108}, 180501 (2012).



\bibitem{Wen_Plaquette}
X.-G. Wen, Phys. Rev. Lett. {\bf 90}, 016803 (2003).

\bibitem{Supp_Mat}
Supplemental Material

\end{thebibliography}

\begin{thebibliography}{2}
\bibitem{Kitaev_2} A. Kitaev, Ann. Phys. {\bf 321}, 2 (2006).
\end{thebibliography}
\end{document}